\documentstyle[preprint,prd,aps,epsf,psfrag,graphicx,latexsym]{revtex}
\begin{document}
\newcommand{\largepspicture}[1]{\centerline{\setlength\epsfxsize{8cm}\epsfbox{#1}}}
\newcommand{\vlargepspicture}[1]{\centerline{\setlength\epsfxsize{13cm}\epsfbox{#1}}}
\newcommand{\smallpspicture}[1]{\centerline{\setlength\epsfxsize{7.5cm}\epsfbox{#1}}}
\newcommand{\tinypspicture}[1]{\centerline{\setlength\epsfxsize{7.4cm}\epsfbox{#1}}}
\newcommand{\ewhxy}[3]{\setlength{\epsfxsize}{#2}
            \setlength{\epsfysize}{#3}\epsfbox[0 20 660 580]{#1}}
\newcommand{\ewxy}[2]{\setlength{\epsfxsize}{#2}\epsfbox[10 30 640  590]{#1}}
\newcommand{\ewxynarrow}[2]{\setlength{\epsfxsize}{#2}\epsfbox[10 30 560 590]{#1}}
\newcommand{\ewxyvnarrow}[2]{\setlength{\epsfxsize}{#2}\epsfbox[10 30 520 590]{#1}}
\newcommand{\ewxywide}[2]{\setlength{\epsfxsize}{#2}\epsfbox[0 20 380 590]{#1}}
\newcommand{\err}[2]{\raisebox{0.08em}{\scriptsize{$\hspace{-0.2em}\begin{array}{@{}l@{}}
                     \plus\makebox[0.55em][r]{#1}\\[-0.05em]
                     \minus\makebox[0.55em][r]{#2}
                     \end{array}$}}}
\def\k{\kappa}
\def\b{\beta}
\newcommand{\errr}[2]{\raisebox{0.08em}{\scriptsize {$\hspace{-0.3em}\begin{array}{@{}l@{}}\plus\makebox[1.2em][r]{#1}\\[-0.0em]\minus\makebox[1.2em][r]{#2}\end{array}$}}}
\newcommand{\plus}{\makebox[15pt][c]{$+$}}
\newcommand{\minus}{\makebox[15pt][c]{$-$}}
\newcommand{\mpr}{\frac{m_{\pi}}{m_{\rho}}}

\newcommand{\fs}{\; .}
\newcommand{\eq}{eq.~}
\newcommand{\beq}{\begin{equation}}
\newcommand{\eeq}{\end{equation}}
\newcommand{\bea}{\begin{eqnarray}}
\newcommand{\eea}{\end{eqnarray}}
\newcommand{\nn}{\nonumber}

\newcommand{\oks}{{1 \over \kappa_{\rm sea}}}
\newcommand{\okl}{{1 \over \kappa_{\rm sea}^{\rm light}}}
\newcommand{\okss}{{1 \over \kappa_{\rm sea}^{\rm strange}}}
\newcommand{\okst}{{1 \over \kappa^{\rm strange}}}
\newcommand{\okv}{{1 \over \kappa_{\rm val}}}
\newcommand{\okvc}{{1 \over \kappa_{\rm val}^c}}
\newcommand{\okvl}{{1 \over \kappa_{\rm val}^{\rm light}}}
\newcommand{\okc}{{1 \over \kappa_{\rm sea}^c}}

\newcommand{\kss}{\kappa_{\rm sea}^{\rm strange}}
\newcommand{\kst}{\kappa^{\rm strange}}
\newcommand{\ks}{\kappa_{\rm sea}}
\newcommand{\kv}{\kappa_{\rm val}}
\newcommand{\kvc}{\kappa_{\rm val}^c}
\newcommand{\kvl}{\kappa_{\rm val}^{\rm light}}
\newcommand{\ksc}{\kappa_{\rm sea}^c}
\newcommand{\kc}{\kappa_{\rm sea}^c}
\newcommand{\ksl}{\kappa_{\rm sea}^{\rm light}}
\newcommand{\kl}{\kappa_{\rm sea}^{\rm light}}

\newcommand{\mq}{m}
\newcommand{\mqv}{m_{\rm val}}
\newcommand{\mqs}{m_{\rm sea}}
\newcommand{\mql}{m^{\rm light}}
\newcommand{\mqst}{m_{\rm strange}}

\newcommand{\mot}{m_{\sf ss}}
\newcommand{\mtf}{m_{\sf vv}}
\newcommand{\moth}{m_{\sf sv}}

\newcommand{\mps}{m_{\rm PS}^2}
\newcommand{\mpsot}{m_{\rm PS, {\sf ss}}^2}
\newcommand{\mpstf}{m_{\rm PS, {\sf vv}}^2}
\newcommand{\mpsoth}{m_{\rm PS, {\sf sv}}^2}

\newcommand{\mv}{m_{\rm V}}
\newcommand{\mvot}{m_{\rm V, {\sf ss}}}
\newcommand{\mvtf}{m_{\rm V, {\sf vv}}}
\newcommand{\mvoth}{m_{\rm V, {\sf sv}}}

\newcommand{\mn}{m_{\rm N}}
\newcommand{\mnot}{m_{\rm N, {\sf ss}}}
\newcommand{\mntf}{m_{\rm N, {\sf vv}}}
\newcommand{\mnoth}{m_{\rm N, {\sf sv}}}

\newcommand{\md}{m_{\Delta}}
\newcommand{\mdot}{m_{\Delta, {\sf ss}}}
\newcommand{\mdtf}{m_{\Delta, {\sf vv}}}
\newcommand{\mdoth}{m_{\Delta, {\sf sv}}}

\newcommand{\chisq}{\chi^2/{\sf d.o.f}}

\newcommand{\tkl}{{{\small\sf T}{$\chi$}{\small\sf L}} }
\newcommand{\tkll}{{{\sf T}{$\chi$}{\sf L}}}
\newcommand{\sesaml}{{\sf SESAM}}
\newcommand{\sesam}{{\small\sf SESAM }}
\newcommand{\sesams}{{\small\sf SESAM}'s }
\newcommand{\sesamc}{{\small\sf SESAM} collaboration}

%
\renewcommand{\baselinestretch}{1.25}
\preprint{HLRZ 1998\_25, WUB 98-21} 

\draft
\title
{Light and Strange Hadron Spectroscopy \\
with Dynamical Wilson Fermions}

\author{N.~Eicker, P.~Lacock,   K.~Schilling, A.~Spitz}

\address{HLRZ, Forschungszentrum J\"ulich, 52425 J\"ulich and\\
DESY, 22603 Hamburg, Germany}

\author{U.~Gl\"assner, S.~G\"usken, H.~Hoeber, Th.~Lippert,\\
  Th.~Struckmann, P.~Ueberholz,  J.~Viehoff}
\address{Fachbereich Physik, Bergische Universit\"at, Gesamthochschule
Wuppertal\\Gau\ss{}stra\ss{}e 20, 42097 Wuppertal, Germany}
\author{ G.~Ritzenh\"ofer}
\address{Center for Theoretical Physics, Laboratory for Nuclear
  Science
and Department of Physics, Massachusetts Institute of Technology,
Cambridge, MA 02139}
\author{\sesamc}


\maketitle


\tighten
\begin{abstract}
We present the final analysis of the light and strange hadron spectra
 from a full QCD lattice simulation with two degenerate dynamical sea
 quark flavours at $\beta = 5.6$ on a $16^3 \times 32$ lattice.  Four
 sets of sea quark masses corresponding to the range $.69
 \leq m_\pi/m_\rho \leq .83$ are investigated.
  For reference we also ran a quenched simulation
 at $\beta_{\sf eff} = 6.0$, which is the point of equal lattice
 spacing, $a_{\rho}^{-1}$.

In the light sector, we find the chiral extrapolation to physical u-
and d- masses to present a major source of uncertainty, comparable to
the expected size of unquenching effects. From linear and quadratic
fits we can estimate the errors on the hadron
masses made from light quarks to be on a 15 \% level prior
to the continuum extrapolation.  For the hadrons with strange
valence quark content, the $N_F = 2$ approximation to QCD appears not
to cure the well-known failure of quenched QCD to reproduce the
physical $K-K^*$ splitting.

\end{abstract}
\newpage
\clearpage
\section{Introduction}

The nonperturbative computation of hadronic properties from Quantum
Chromodynamics (QCD) presents a major challenge in the unraveling of
quark flavourdynamics from hadronic experiments. The methods and tools
of lattice gauge theory have been refined over the past two decades
resulting in rather precise results (to the level of a few percent
accuracy in the physical spectrum of light hadrons, i.e. after chiral and
continuum extrapolations) {\it within the quenched
approximation}\cite{yoshie}. High statistics quenched lattice 
studies on large lattice volumes  revealed that
the effects of dynamical fermions on spectrum and matrix elements
appear to lie   within a 10  to 20 \% range\cite{weingarten,tsukuba}.

  The `solution' of the full QCD binding problem with lattice methods,
on the other hand, is still very much lagging behind.  This is mainly
due to the high cost in compute effort to encompass the fermionic
determinant in the underlying stochastic sampling procedures.  The
simulation of large lattices in full QCD is definitely a task that
requires the power of the upcoming teracomputers. Nevertheless, with
the computing power of some several hundred of teraflops hours it is of
considerable interest to tackle QCD vacuum polarization effects by
looking -- on intermediate volumes in the scaling regime -- at
quantities with inherent sea quark dependence such as the $\pi $N
$\sigma$-term, the $\eta '$-mass, and the quark spin content of the
nucleon.

A full QCD simulation with Wilson fermions is particularly expensive,
as the fermionic operator in this case carries more degrees of freedom
than in the staggered formulation, and its  chiral extrapolation is
more cumbersome as the chiral point fluctuates with the gauge field on
a finite system.  \sesam is a second generation simulation which
is still exploratory, using  Hybrid Monte Carlo (HMC)\cite{hmc} at $\beta
= 5.6$ on $16^3\times 32$ lattices.  For technical reasons, we work
with two (degenerate) dynamical fermions, $N_f = 2$. We devised
several improvements in order to accelerate the computation of the
fermionic force\cite{ourmethods}.  In this way, on the available
APE100 hardware\cite{ape}, we could achieve HMC histories of sufficient
lengths for a safe estimate of autocorrelation times. This provides a
sound basis for the error analysis.

In a full QCD computation there is no difference between sea quarks,
which contribute to the fermion determinant, and 
valence quarks, which occur in the hadron operators that are employed 
to excite hadronic states from the QCD vacuum.

In our $N_f = 2$ scenario, however, one is forced to introduce {\it
  `valence' quarks different from sea quarks}, as soon as one wishes
to deal with hadrons carrying strangeness. In a recent
letter\cite{qmasses} devoted to the determination of the light and
strange quark masses, we have therefore considered hadronic
correlators on a set of three different sea quark masses, with valence
quark content {\it both equal and different} to that of the underlying
sea quark and presented a consistent approach to analyse such
`semiquenched' data.

In this paper we will extend that  work {\it from three to four}
different sea quark masses and present a detailed study of the light
and strange hadron spectra. We shall
identify sea and valence quarks in the light sector (of u and d quarks)
and resort to the semiquenched ansatz with respect to the strange
quarks, as living in a sea of light quarks.  

For reference, we perform a {\it concomitant quenched simulation} on
equal lattice spacing and volume, at $\beta_{eff} = 6.0$\footnote{This
  value is at the onset of the (quenched) scaling regime.}.  While
unquenching definitely leads to a considerable decrease of the light
quark mass estimate\cite{qmasses}, we find -- within our errors -- no
visible sea quark effects both on the light and strange hadron masses.
%
\section{Simulation Details \label{simulation_details}}
%
%
\subsection{Hybrid Monte Carlo \label{hybrid}}
%
We have performed a large scale simulation of full QCD at $\beta =
5.6$ with two degenerate flavors of dynamical Wilson fermions. We have
generated lattices of extent $16^3 \times 32$ at four different values
of the sea-quark hopping parameter using the $\Phi$-version
\cite{GOTTLIEB} of the HMC algorithm.  The parameters used in the HMC
update and the statistics for the complete runs on the 256 node
APE100/Quadrics QH2 are given in table \ref{simulation}.

The CPU costs of the HMC are mostly due to the time consuming repeated
solution of the linear system $M^{\dagger}\,M\,X=\Phi$ with $M$ being
the Wilson fermion matrix. Throughout our simulation we employed the
bi-conjugate gradient stabilized algorithm (BiCGStab) which has been
demonstrated to be the most efficient Krylov sub-space solver for
Wilson fermion inversions \cite{ourmethods}.  Using BiCGStab, we
computed the linear system in a two step procedure
\begin{equation}
M^{\dagger}\,Y = \Phi,\qquad M\,X = Y.
\end{equation}

In the first stage of the simulation, we preconditioned by use of the
o/e decomposition of the Wilson fermion matrix $M\rightarrow M_e$
\cite{ROSSI}, $M_e={\bf 1}-\kappa^2\,D_{eo}D_{oe}$ referred to as o/e
in table \ref{simulation}.
%
%
In a later stage of the simulation we switched from the thinned o/e
representation $\det(M_e)$ to the full fermion determinant $\det(M)$
in order to employ the locally lexicographic SSOR preconditioner
\cite{FISCHER} which has been shown to offer up to a factor of 2 less
computational costs than o/e preconditioning.  In table \ref{simulation} we
refer to this part of the simulation as `SSOR'.

As a third improvement of the molecular dynamics part within our HMC, we
have implemented the chronological start vector guess \cite{BROWER}.
The optimal depth of the extrapolation, $N_{CSG}$, has been determined
empirically for each $\ks$ and with respect to the representation of
the fermionic determinant as listed in table \ref{simulation}.

We have selected the time step size and the number of molecular
dynamics steps, $N_{md}$, to yield an acceptance rate of $>70 \%$ in
the global Monte Carlo decision of HMC.  With decreasing sea quark
mass we can observe a variation of the acceptance rate from $85 \%$ to
$73 \%$.  We have varied the trajectory length $N_{md}$ by numbers
uniformly distributed in the range $\pm 2\sqrt{N_{md}}$ as recommended
in Ref.~\cite{MACKENZIE} to avoid deadlocks in periodic orbits of
phase space due to the presence of well defined Fourier modes.
%
%

The chosen stopping accuracy, $R$, of the iterative solution of
$M^{\dagger}\,M\,X=\Phi$ is the only source of systematic error of the
HMC. We have defined the convergence criterion by
$R=\frac{||M\,X-\Phi||}{||X||}=10^{-8}$ throughout our simulations,
working at the level of APE's 32-bit precision. Beyond $R<10^{-7}$,
the difference Hamiltonian $\Delta H$ for the global Monte Carlo
decision, computed in double precision, does not vary significantly.

We proceeded adiabatically from large to small sea quark masses
and, after thermalizing for more than 500 trajectories at each $\ks$,
for each sea quark mass, we have generated 5000 trajectories.  From
these correlated samples we have chosen 200 decorrelated lattices per
sea quark mass, see table \ref{simulation}.
%
\subsection{Error estimates \label{error}}
%
Since HMC is a Markov process, one is faced with the problem of
autocorrelation of the generated series of trajectories. Of course one
would like to aim at a decorrelated sample of configurations. However,
since the generation of
trajectories for full QCD is extremely costly we cannot afford to skip
many trajectories as one can do in quenched simulations.  In order to
control the statistical quality of the measured signals we have to
carefully study the autocorrelation of the Markov chain.

We paid attention to keep stable conditions for the HMC dynamics to
evolve rather than re-tuning HMC parameters during production. This
provides the setting for a reliable determination of the
autocorrelation times related to various hadronic quantities.

For all four $\ks$ values, both exponential and integrated autocorrelation
times of various gluonic and fermionic observables have been measured.
The relevant quantity for the error determination is the integrated
autocorrelation time $\tau_{\rm int}$. We found $\tau_{\rm int}$,
which is observable-dependent, to be bound from above by $\tau_{\rm
  int}^{\Lambda}$ of the smallest eigenvalue $\Lambda$ of the fermion
matrix\footnote{In ref.\cite{SESauto} we shall present a detailed
  account of the underlying auto-correlation analysis, and we shall
  propose a scaling rule for the critical slowing down of the HMC.}.
$\tau_{\rm int}^{\Lambda}$ varies between $15$ for $\ks=0.156$ and
$30$ for $\ks=0.1575$, however, the integrated autocorrelation times
of most hadronic observables lie well below this limit. Therefore, we
have decided to analyze every 25th trajectory for spectrum and decay
constants, after thermalization.

\begin{figure}[htb]
\begin{center}
\noindent\parbox{16.0cm}{
\parbox{7.0cm}{\epsfxsize=7.0cm\epsfbox{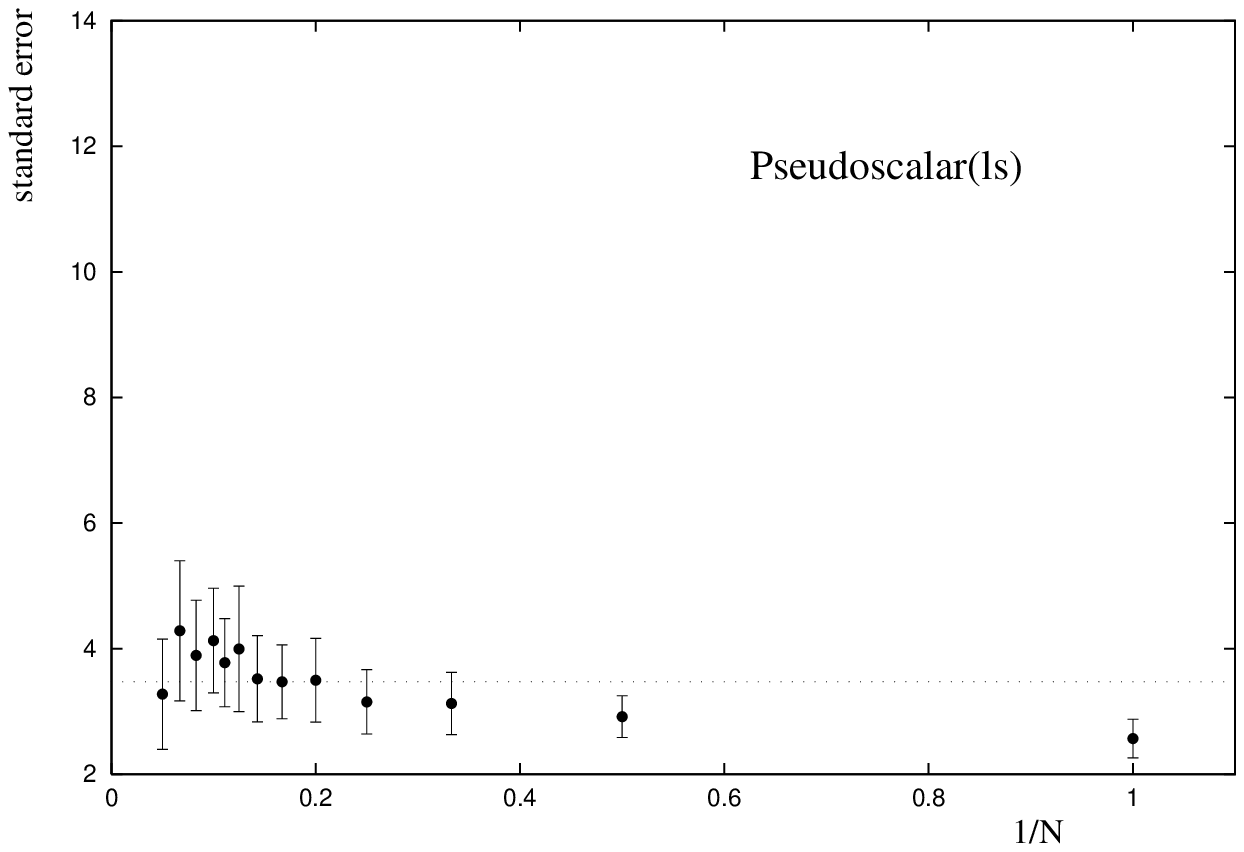}}
\parbox{7.0cm}{\epsfxsize=7.0cm\epsfbox{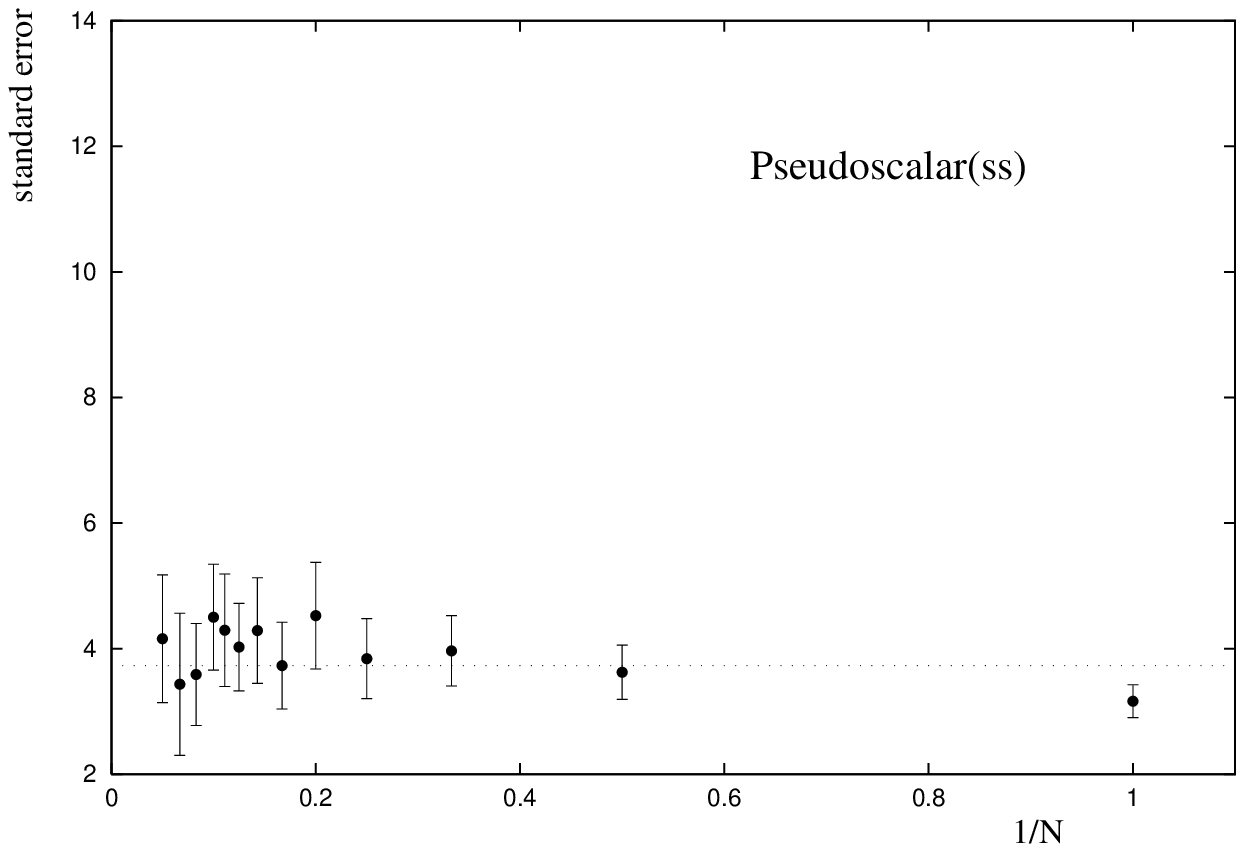}}
\\ \linebreak
\parbox{7.0cm}{\epsfxsize=7.0cm\epsfbox{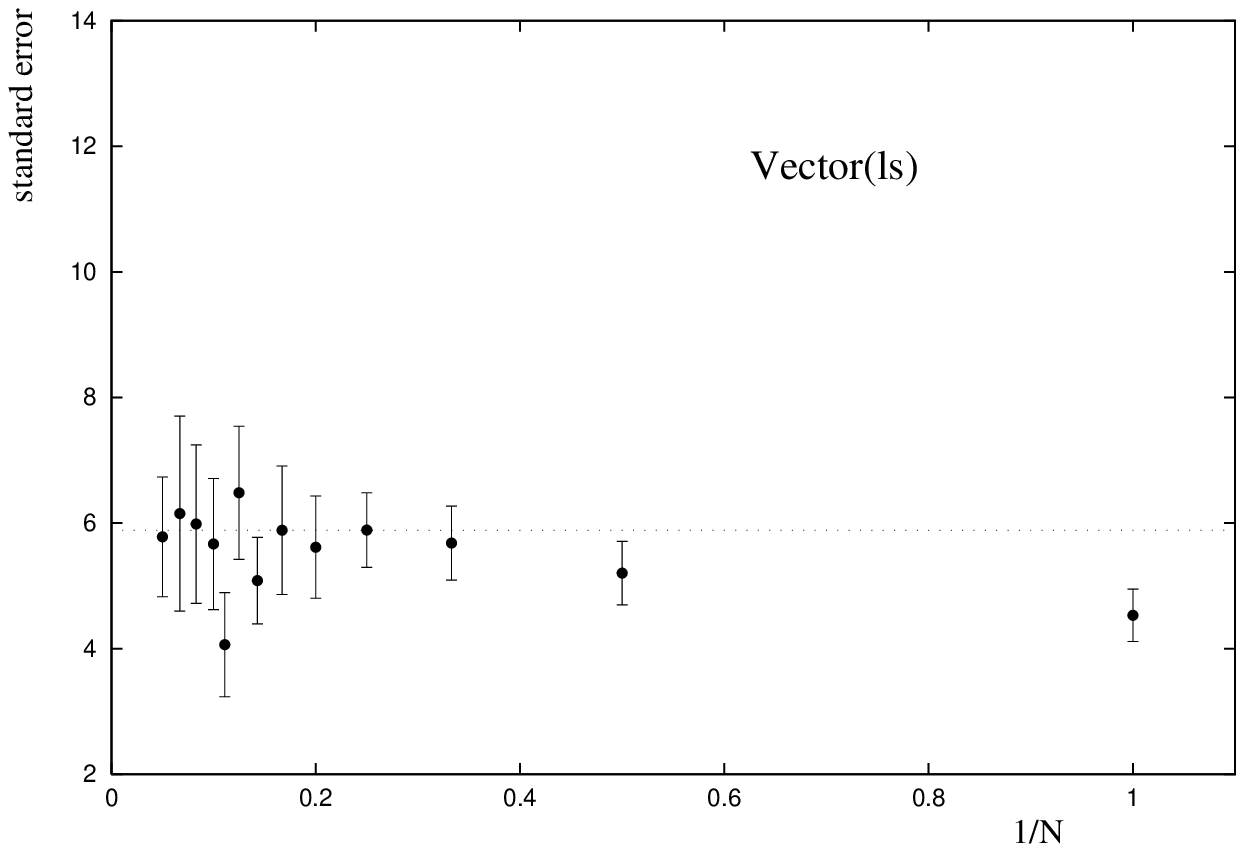}} 
\parbox{7.0cm}{\epsfxsize=7.0cm\epsfbox{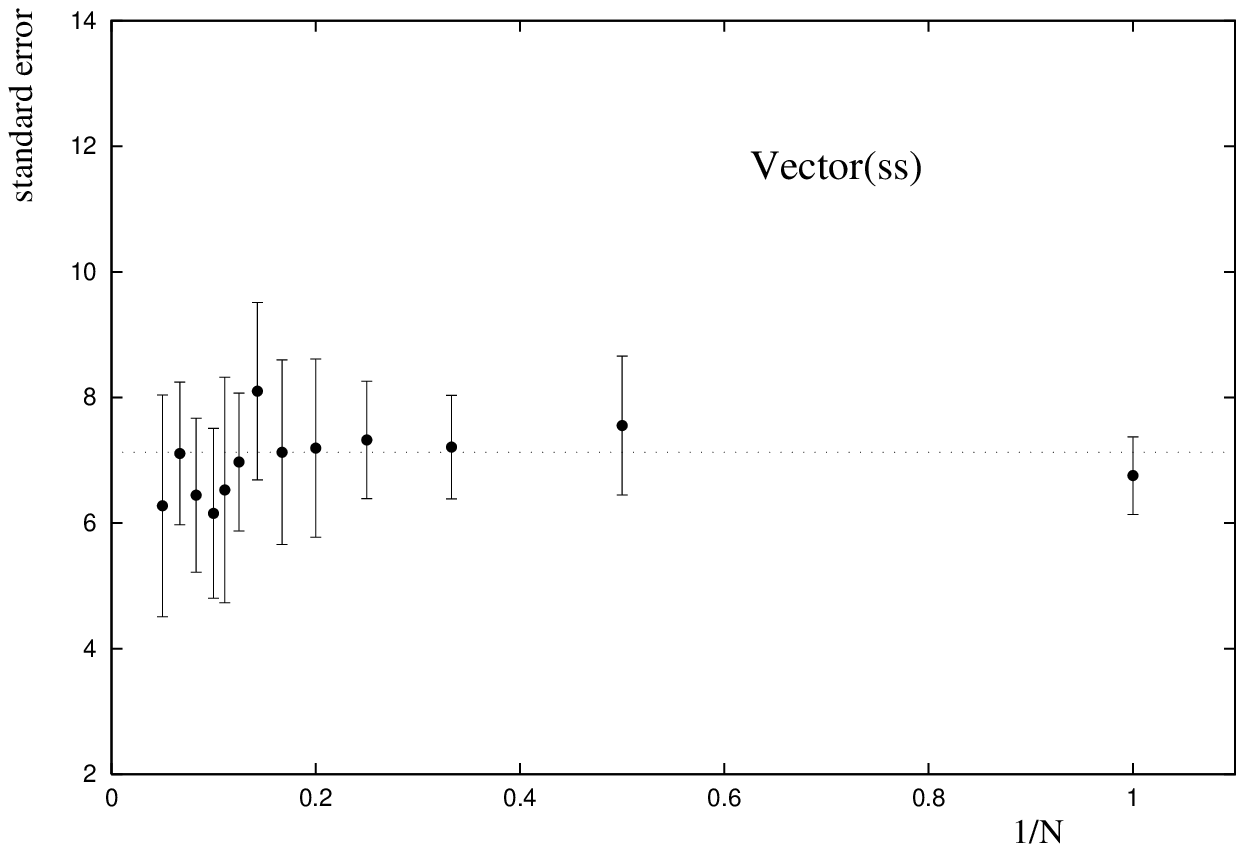}}\\
 \linebreak
\parbox{7.0cm}{\epsfxsize=7.0cm\epsfbox{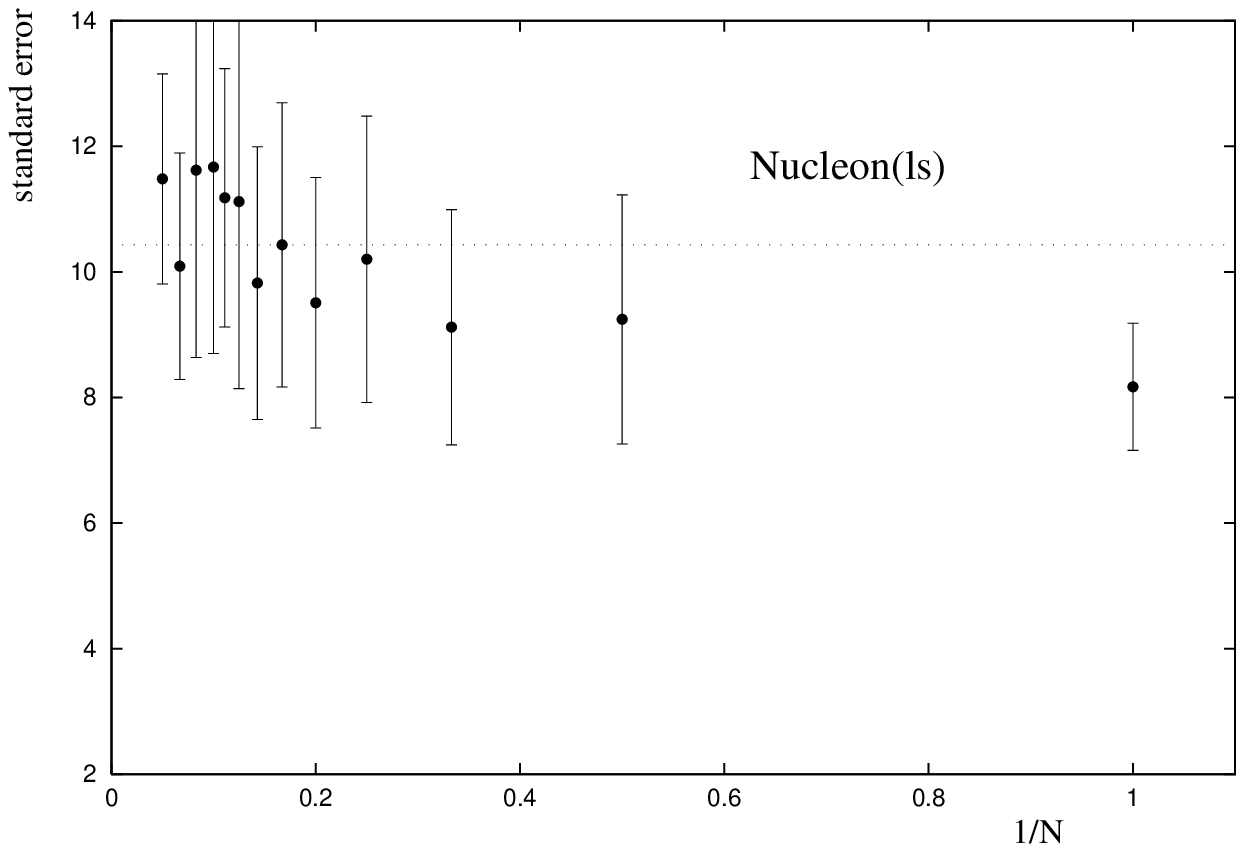}}
\parbox{7.0cm}{\epsfxsize=7.0cm\epsfbox{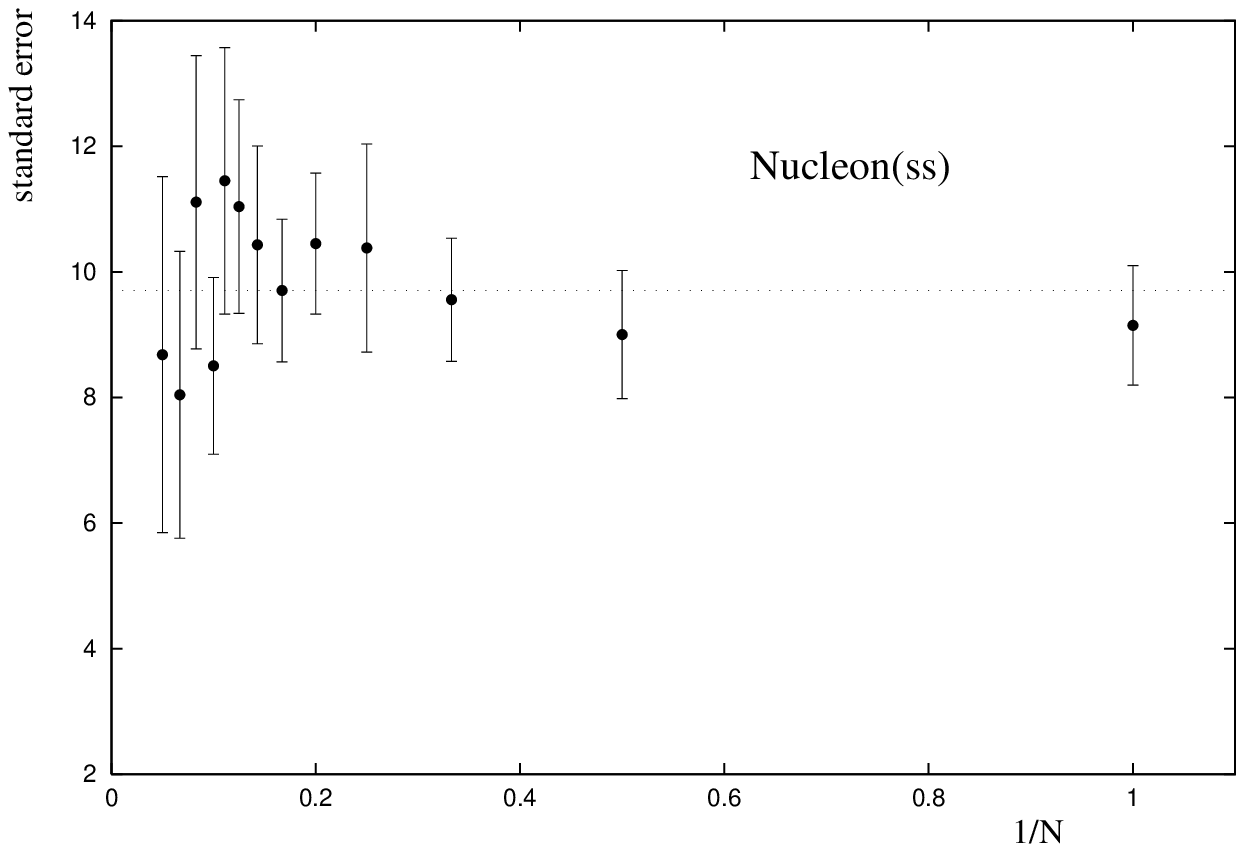}}\\
 }
\caption{\label{fig_sigsig} The standard error of the masses of 
pseudoscalar, vector and nucleon as a function
 of the inverse block size}
\end{center}
\end{figure}

In order to account for possibly remaining correlations within our
hadronic observables we have carried out a blocking investigation. For
our smallest sea quark mass we show in fig.\ref{fig_sigsig} the errors
of $m_\pi$, $m_\rho$ and $m_N$, as a function of the blocking size.
At block size 4 to 6 we find the jackknife errors to run into
plateaus.  Accordingly, we shall use a block size of six throughout our
analysis applying the bootstrap procedure. Errors (on the blocked data) are obtained from bootstrap samples with 250 entries each.  A
similar analysis of our quenched data shows no increase in error with
the block size (quenched configurations are generated with an
over-relaxed Cabbibo-Marinari heatbath update and are separated by 250
sweeps).

We remark that we have investigated the decorrelation efficiency of
the HMC with respect to topology on the chosen samples\footnote{This
  investigation is a prerequisite for the
  investigation of quantities related to topology.} \cite{TOPOL}.
%
%
It is gratifying that we could establish sufficient tunneling of the
topological charge through the topological sectors for the four $\ks$
values investigated. For our smallest quark mass we determined an
integrated autocorrelation time with respect to the topological charge
of $\tau_{\rm int}\approx 50$.  Furthermore, we analyzed some hadronic
quantities---which do not explicitly depend on topological
effects---according to the topological charge content of the
configurations. The result of this analysis is that no
significant dependence on the
topological sector was found.
%
%
\subsection{Hadronic observables \label{operators} }
%
At each of our four sea quark values we have investigated
zero momentum two-point functions, 
\beq
C_{AB}(t) = \sum_{\vec x}  \langle 0 | 
\chi_A^{\dagger}(x) \chi_B(0) | 0 \rangle \; ,
\end{equation}
with hadronic excitation operators $\chi$ as listed in table
\ref{ops}. We combined light-quark propagators with hopping parameters
equal and different to that of the underlying sea quark, thus
providing ourselves with fifteen hadronic mass combinations  at the two heaviest
sea quarks and ten at the two lightest (see table \ref{kappas1} for
the complete list).

We use the gauge-invariant Wuppertal-smearing procedure
\cite{Wupsmear} to calculate ``smeared-local'' ({\it sl}) and
``smeared-smeared'' ({\it ss}) correlators. The smearing parameter is
chosen to be $\alpha =4$, with $N=50$ iteration steps. In an attempt
to further improve on our ground state projection we carried out an
additional run with $100$ smearing iterations at $\ks = 0.1565$;
although this rendered a somewhat faster drop into the ground state it
did not alter our fit results. Plots with $\ks = 0.1565$ are from our
run with $N=100$.\par

Our analyses are based on global masses as extracted from
single-exponential fits to the correlators:
\begin{eqnarray}
\label{single_ex}
C(t)_{\rm{mes}} & = & A ( e^{-m t} + e^{-m (T-t)} ) \nonumber \\
C(t)_{\rm{bar}} & = & A e^{-m t} \, ,
\end{eqnarray}
with $T=32$.
As a cross check, we also determined effective local masses.  For
mesons they are computed  iteratively from the implicit equation
\begin{equation}
\frac{C_{AB}(t)}{C_{AB}(t+1)} 
= \frac{e^{-m_{\rm{eff}}(t) t} + e^{-m_{\rm{eff}}(t) (T-t)}}
       {e^{-m_{\rm{eff}}(t)(t+1)} + e^{-m_{\rm{eff}}(t) (T-t-1)}}
       \quad ,
\end{equation}
while for baryons they are determined in the standard manner from the
plateau of local masses:
\begin{equation}
m_{\rm{eff}}(t) = \rm{log} \frac{C_{AB}(t)}{C_{AB}(t+1)}\; .
\end{equation}
We use the smeared-smeared data to obtain both masses and
amplitudes. The fit ranges are determined by keeping the upper limit
fixed half-way across the lattice , while the lower cut in $t$ is
varied in the interval $7-10$\footnote{The smeared-local data yield
consistent results but correspond to smaller fit ranges in $t$.}.  The
mass plateau range with the best $\chisq$-value is selected as fit
interval. 

\begin{figure}
\begin{center}
\epsfxsize=14cm\epsfbox{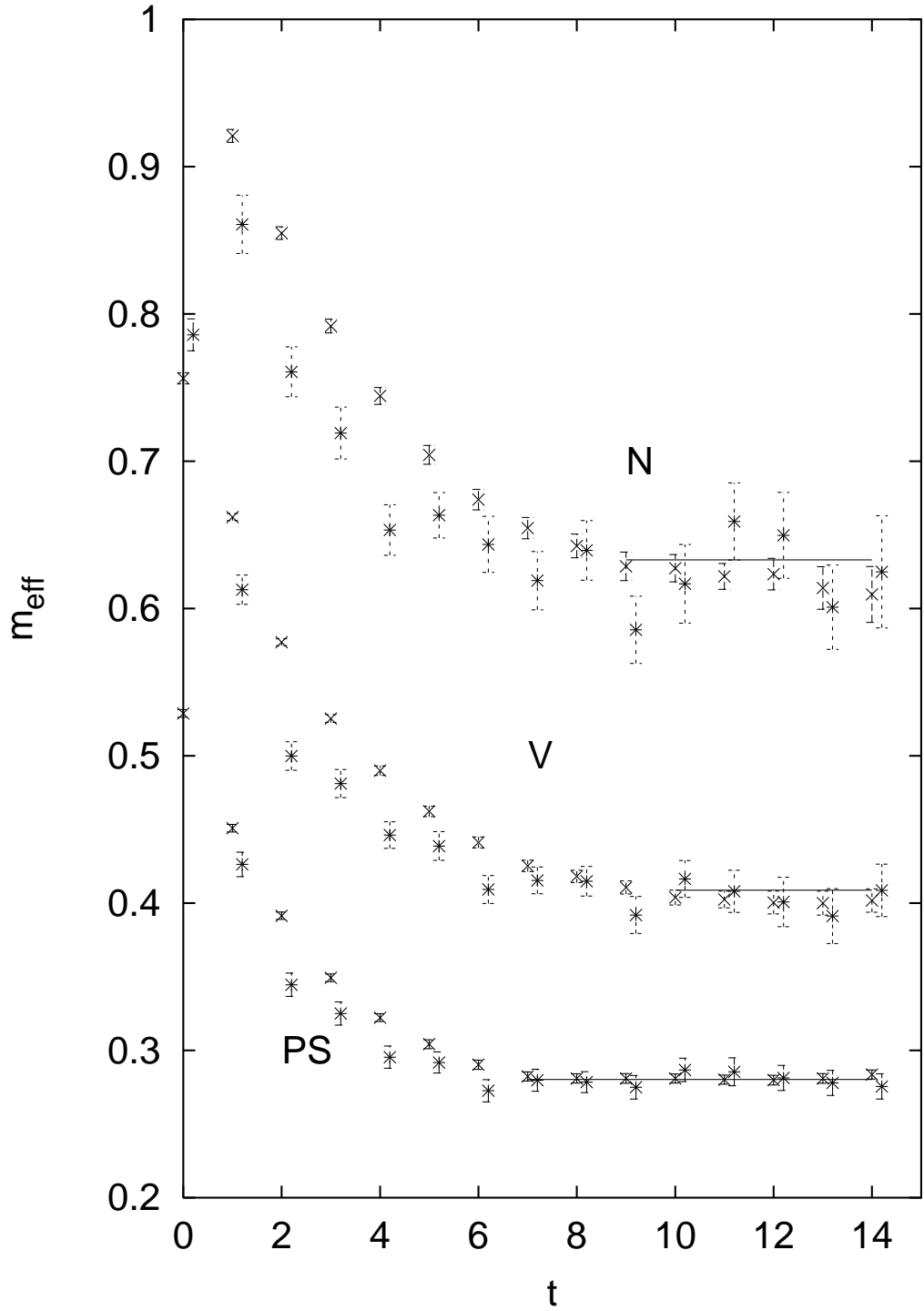}
\caption{Effective masses at
$\kappa_{\rm{sea}}=0.1575$. $\ast$ are smeared-smeared and $\times$ are
smeared-local data. The results of our fits to
the smeared-smeared correlators (not
to the effective masses!) are indicated by solid lines.\label{fig_mass_eff}}
\end{center}
\end{figure}

Figure \ref{fig_mass_eff} illustrates the quality of our data by showing
the different effective local masses in comparison to the global
masses from correlated fits to the two-point functions, in the
optimal fit ranges. We find that uncorrelated fits lead to consistent
results.  \par For future reference our `raw data' from these mass
fits are collected in tables \ref{raw_mass_156} to
\ref{raw_mass_1575}. By inspection of these tables we
retrieve $m_\pi/m_\rho$ ratios of $0.833(5)$, $0.809(15)$, $0.758(11)$
and $0.686(11)$ at $\kappa_{\sf sea} = 0.1560, 0.1565, 0.1570$ and 0.1575
respectively.

We determine the pseudoscalar and vector decay constants from the
respective current matrix elements on the lattice: 

\beq f_\pi = {1\over m_\pi} Z_A \langle 0 | A_0^l | \pi \rangle \, , \hspace{1cm}
{\epsilon_\mu \over f_V} = {Z_V \over m_V^2} \langle 0 | V_\mu^l | V
\rangle .
\label{dec}
\eeq

  The matrix elements are extracted from a direct fit to the ratios
\begin{eqnarray}
R_{PS} & = & \frac{\langle 0 | A_0^l A_0^s | 0 \rangle}
               {\langle 0 | A_0^s A_0^s | 0 \rangle^{1/2}}
= \langle 0 | A_0^l | \pi \rangle e^{-m_{PS} {T \over 4}} {1 \over
               \sqrt{2 m_{PS}}} {\sf cosh}^{{1 \over 2}}\left( m_{PS} (t -
               { T \over 2}) \right) \nn \\
R_V &=&  \frac{\langle 0 | V^l V^s | 0 \rangle}
         {\langle 0 | V^s V^s | 0 \rangle^{1/2}} 
= 3^{-1/2}\langle 0 | V_\mu^l | V \rangle e^{-m_V {T \over 4}} {1 \over
               \sqrt{2 m_V}} {\sf cosh}^{{1 \over 2}}\left( m_V (t -
               { T \over 2}) \right) \, , \label{decay_fit} 
\end{eqnarray}
where the superscripts $l$ and $s$ denote local and smeared
operators, respectively. Note that in the second equation
the operators $VV$ stand generically for $\sum_{k=1}^{3} V_kV_k$ while
the operator $V$ on the {\it r.h.s.} denotes $\sum_{k=1}^{3} V_k$.

The masses in equations \ref{dec} and
\ref{decay_fit} are fixed to the values obtained from the mass fits
(given in tables \ref{raw_mass_156} through \ref{raw_mass_1575}).  

\par The `raw data' for the lattice matrix elements $\langle 0 | A_0^l
| \pi \rangle$ and $3^{-1/2}\langle 0 | V^l | V \rangle$ as well as
$f_\pi/Z_A$ and $1/(f_V Z_V)$ are collected  in tables
\ref{raw_decay_156} to \ref{raw_decay_1575}.
 \par 
The renormalization factors $Z_A$ and $Z_V$ are computed
perturbatively, as explained in the Appendix.

%
\section{Chiral Extrapolations}
\subsection{The light sector \label{s_light}}
%
We will first present our results for particles and decay constants
containing nonstrange quarks only. It is obvious to identify the degenerate
sea quarks in our simulation with the $u$ and $d$ quarks since,
naively, we expect the lightest sea quarks to make the largest effect
on the hadronic properties.  In this scenario the light hadrons are
determined from our raw data by a chiral extrapolation in quark mass.
We call this setting ``symmetric'', since it involves data points with
equal sea and valence quark masses only.
 
At this stage one should remember that full QCD vacuum configurations
on different sea quark sectors are manifestly decorrelated. This has
some bearing on the error analysis of hadron spectra, differently from
the quenched situation where one normally determines entire hadron
mass trajectories configuration wise, with ensuing point to point
correlations.  It will be interesting to trace the impact of this
peculiarity on the accuracy of hadron masses and decay amplitudes
under chiral extrapolation, in the full QCD situation.

%
\subsubsection{Masses and decay constants\label{ssmasses}}
%
The pseudoscalar mass is used to extract the critical hopping
parameter $\kc$ while the value of the light hopping parameter $\kl$
is set by the condition\footnote{In this paper we use the convention
that physical masses\cite{PRD} are written in capital letters, while
lattice masses are denoted by small letters.}: \beq {m_{\rm PS}(\kl)
\over \mv(\kl)} = {M_{\pi} \over M_{\rho}} = 0.1785 \, .
\label{ratiolight}
\eeq
 The isospin symmetric bare light quark mass is given by:  
\beq
\mql = {1 \over 2} (\okl - \okc) \; .
\eeq 
\par
\begin{figure}
\epsfxsize=14cm\epsfbox{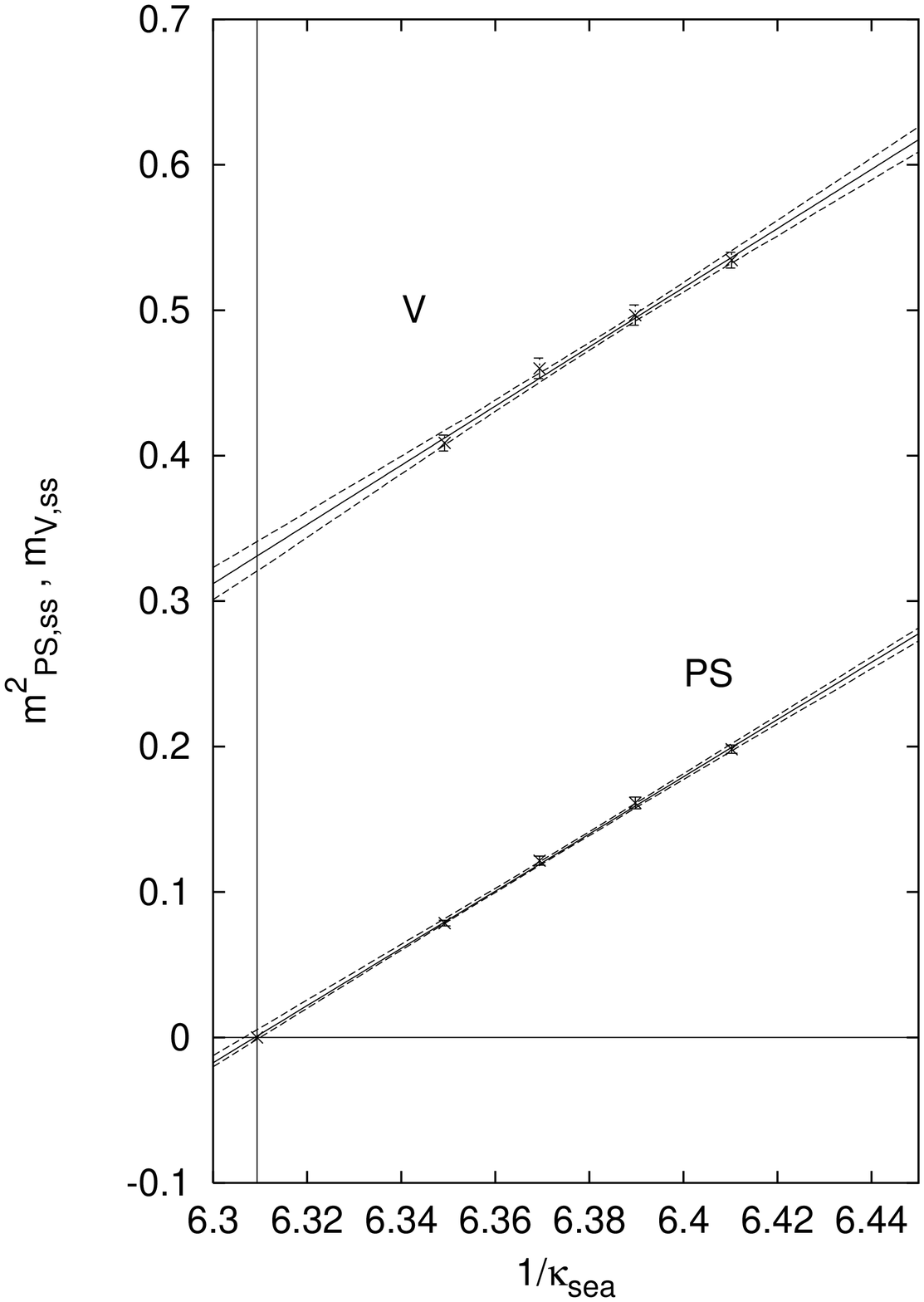}
\caption{$\mpsot$ and $\mvot$ as a function of $\oks$ (in lattice units).\label{fig_symmetric_mesons}}
\end{figure}

Linear fits to our data for the pseudoscalar ($\mpsot$) and vector
masses ($\mvot$) with $\kappa_{\rm sea} = \kappa_{\rm val}$ are shown
in fig.\ref{fig_symmetric_mesons}. The resulting parameter values from the
extrapolations are given in table \ref{table_pr}, where we employ the
following notations: \bea
m^2_{PS} &=& a + b { 1 \over \kappa_{\rm sea}} \\
m_V &=& m^{crit}+ c \mqs + e \mqs^2 \quad \mbox{with} \quad m_{sea} = {1
  \over 2} ({1 \over \kappa_{\rm sea}} - \okc) . \label{traj} 
\label{eq_fit_symmetric}
\eea
We find the pseudoscalar data to be well described  by the linear ansatz,
the fit yielding  the critical value of $\kappa_{sea}$ to be 
\beq
\ksc = 0.158507 \err{41}{44} \, ,
\label{eq_kcsym}
\eeq with $\chisq = 0.6$.  For the vector particle, both linear and
quadratic parametrizations yield acceptable fits, with $\chisq =
0.75 $ and $\chisq = 0.27$, respectively. For the light
hopping parameter, we find
 \beq
\label{klight}
 \ksl = 0.158462 \err{41}{42}\;\mbox{(linear)} \quad , \quad \ksl = 0.158471 \err{45}{45}\; \mbox{(quadratic)} \,
.
\label{eq_klight}
\eeq In the light sector   we 
quote the results from the linear ansatz,  using the quadratic fit to
estimate the systematic uncertainties.  This gives the
following value for  the unrenormalized light quark mass \beq \mql =
0.000901(54)(184)\, , \eeq the second error being the systematic
uncertainty. Note that this value is consistent with our previous
estimate, $\mql = 0.00088(6)$ \cite{qmasses},
obtained from simulations on three sea quark masses.

\begin{figure}
\epsfxsize=14cm\epsfbox{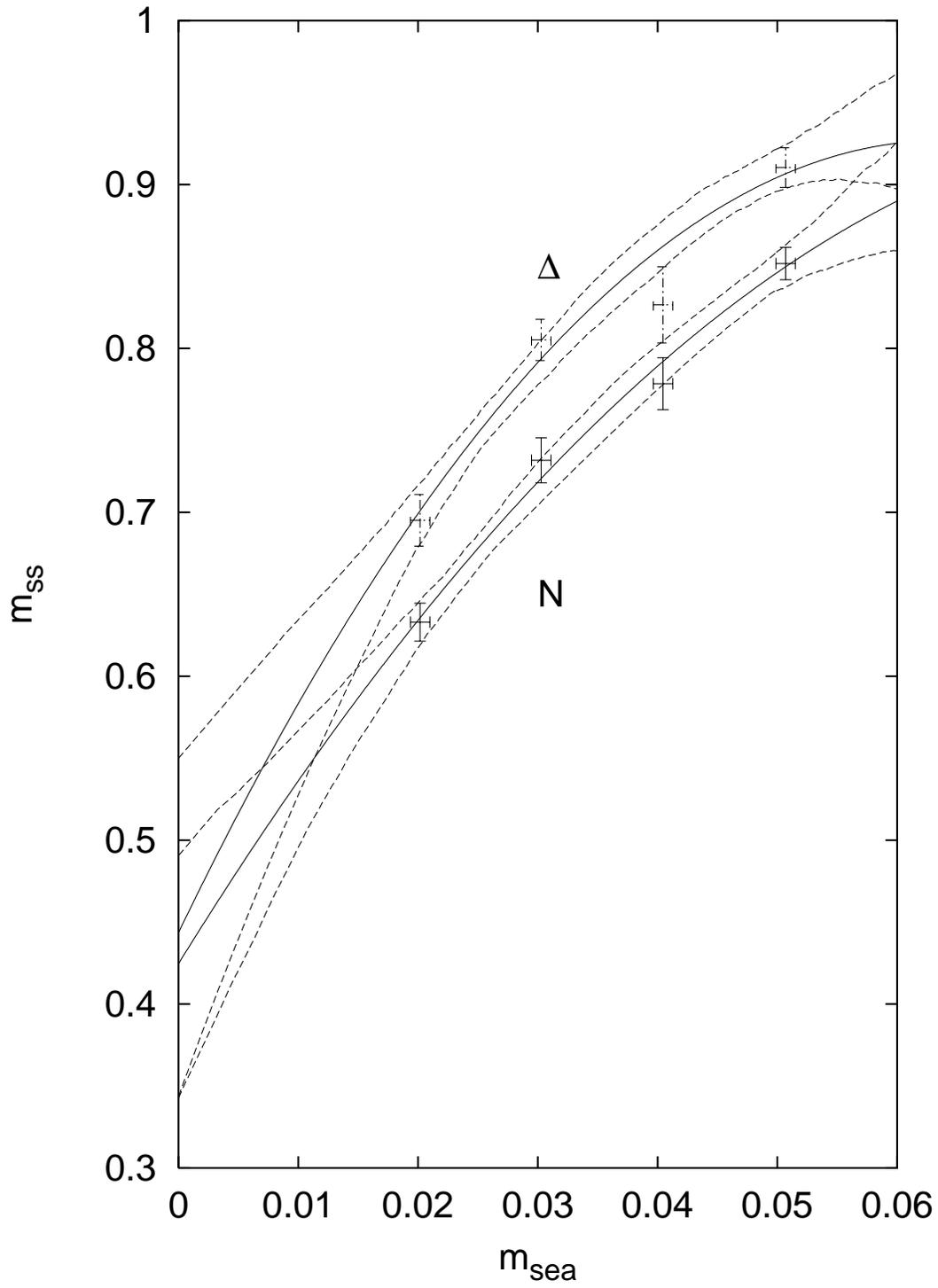}
\caption{$m_N$ and $m_{\Delta}$ as a function of $m_{sea}$ (in lattice units).
\label{fig_symmetric_baryons}}
\end{figure}

We can now predict the nucleon and $\Delta$ masses and the $\pi$ and
$\rho$ decay constants by chiral extrapolation to the point $\ksl$.
The resulting fit parameters, in the notation of of eq. \ref{eq_fit_symmetric},
are collected in table \ref{table_pr}. The extrapolations of the
baryonic masses are visualized in fig.\ref{fig_symmetric_baryons}. It
turns out that their $m_{\sf sea}$-dependence is by a factor 2 to 3
stronger than in the mesonic case, leading to a statistical error on
the mass extrapolations for nucleon and $\Delta$ of 16 and 22 \%,
respectively. By comparing the deviations among  linear and
quadratic extrapolations, see table \ref{table_pr}, we might estimate a
systematic error of 15 \% and 24 \%, respectively, which is covered
by the statistical error, however.  In
order to put these numbers  into perspective, one should be aware that
we are extrapolating down from $m_\pi/m_\rho = .686$ on the basis of raw
data, which carry statistical errors in the range of one to two \%
(see tables \ref{raw_mass_156} to \ref{raw_mass_1575}). Just for
reference: in state-of-the-art quenched
simulations\cite{yoshie}  the Tsukuba group achieves
statistical errors in the region of .5 to 1 \%, in the range of
$m_\pi/m_\rho$ down to a value of $0.4$\cite{tsukuba}!

\begin{figure}
\begin{center}
\noindent\parbox{13cm}{
\parbox{6.5cm}{\epsfxsize=6.5cm\epsfbox{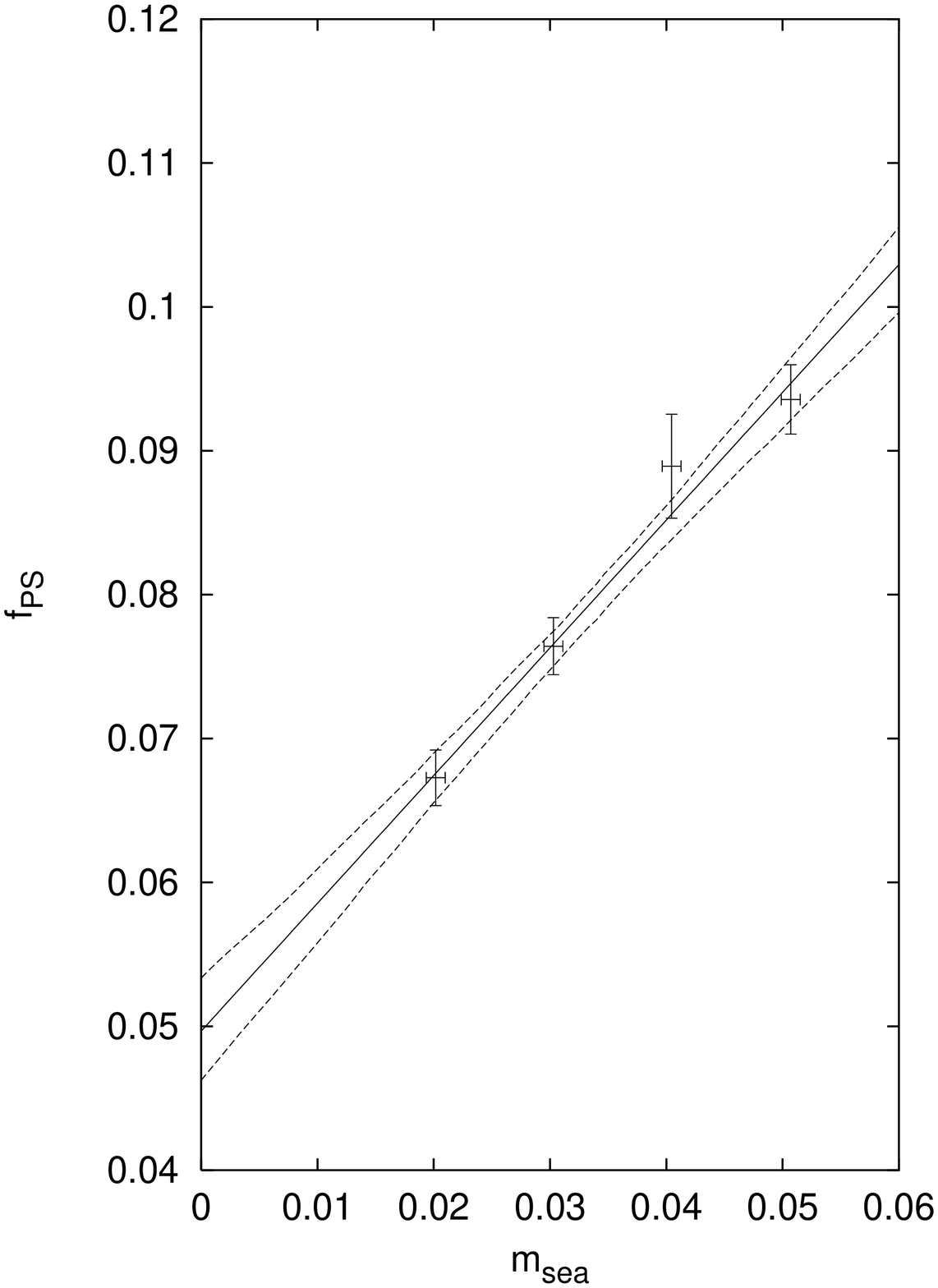}}
\nolinebreak\hfill
\parbox{6.5cm}{\epsfxsize=6.5cm\epsfbox{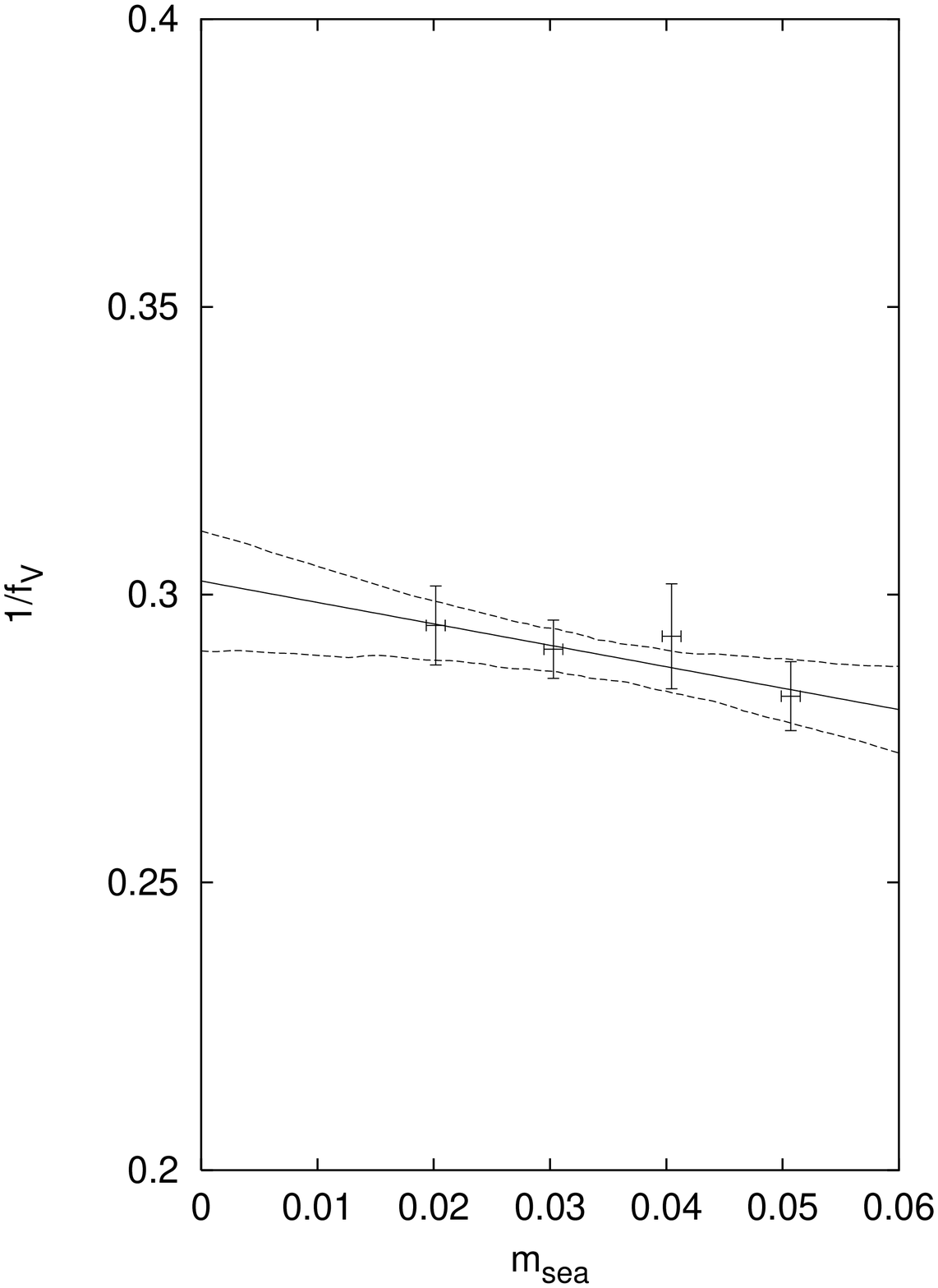}}}
\caption{Linear chiral extrapolations of $f_{PS}$ and $1/f_{V}$
\label{fig_fpi_frho}}
\end{center}
\end{figure}

For the decay constants we proceed similarly. The renormalized data
are displayed in fig. \ref{fig_fpi_frho}; they favour the linear
extrapolation (see also table \ref{table_pr}). Again, by comparing
linear and quadratic results, we estimate our systematic
uncertainties; they amount to 15 and 3 \% for $f_\pi$ and $f_\rho$,
respectively.

%
\subsection{The strange sector}
%

%
So far we have used $M^2_{\pi}$ and ${M_\pi \over M_\rho}$ to set the
values of the hopping parameter values in the chiral limit and 
at the u  quark mass.  In the following we shall briefly
describe our procedure to determine the value of the hopping parameter
related to  the strange quark mass\cite{qmasses}.

Our simulations are based on two `active', degenerate sea quarks,
which we identify with the light quarks.  The strange quark in this
setting has to be treated as an effectively quenched quark that lives
in the sea of the two physical light quarks.  In order to account for
this situation let us, for the sake of clarity, introduce the
generic notation for various types of hadron masses appearing in the course
of our calculations: \vspace{0.3cm}
\par
\begin{enumerate}
\item
$\mot$ - both valence quarks are identical to the sea quark,
\item
$\moth$ - one valence quark coincides with  the sea quark,
\item
$\mtf$ - both  valence quarks differ from the  sea quark.
\end{enumerate}
Note that the `symmetric extrapolations' operate on the data set
 $\mot$ and suffice  to determine both the critical and light hopping
parameter values, as discussed in section \ref{s_light}.
\par
Let us consider the pseudoscalar masses. In a linear parametrization,
the three mass types can be written  in terms of five  slope
parameters, $c$ to $d''$:
\begin{eqnarray}
\mpsot &=& a' \mqs \, , \nn \\
\mpsoth &=& a'' \mqs + b' \mqv  \, , \nn \\
\mpstf &=& a''' \mqs + b'' (m_{\rm val\, 1} + m_{\rm val\, 2}) \, . \nn 
\end{eqnarray}
In  the symmetric situation, $m_{sea} = m_{val\, 1} =
m_{val\, 2}$,  this mapping has to collapse to degeneracy on the {\it
  l.h.s.}
which leads to  constraints on the slopes. As a result one ends up
with the simple  form
\begin{eqnarray}
\left( 
\begin{array}{c} 
\mpsot  \\
\mpsoth \\
\mpstf  
\end{array}
\right) &=&  
\left( 
\begin{array}{cc} 
 a' & 0  \\
a'-b' & b'\\
a'-2b' & 2b' 
\end{array}
\right)   
\left( 
\begin{array}{c} 
\mqs \\
\mqv
\end{array}
\right)   \, 
\label{quarkmatrix}
\end{eqnarray}
which can be used for simultaneous fitting in $m_{sea}$ and $m_{val}$.
In the spirit of our approach, we will
identify the  light quark mass  with $m_{sea}$,
while the strange quark mass is described by  $m_{val}$.  Note that 
through  the degeneracy requirement, 
 we are effectively left with {\it two independent} slopes
only\footnote{In our previous letter we used three independent such
  slopes\cite{qmasses}.}, $a'$ and $b'$.

Valence quark masses in \eq\ref{quarkmatrix} are defined as:
\beq
\mqv = {1 \over 2}\left( {1 \over \kappa_{\rm v}} - \okc \right)\;.
\label{barequarks}
\eeq
In this setting, with the three types of hadron masses we are in the
position to perform `semiquenched extrapolations' where valence
$\kappa $ values with $\kv \neq \ks$ are admitted.  

\par Within the linear ansatz, other hadronic quantities (like masses
and decay constants) can be written in terms of $m_{sea}$ and
$m_{val}$ in the generic form

\begin{eqnarray}
\left( 
\begin{array}{c} 
m_{\sf ss}  \\
m_{\sf sv} \\
m_{\sf vv}  
\end{array}
\right) &=&  m'^{\rm crit} +  
\left( 
\begin{array}{cc} 
 c' & 0  \\
 c'- d' & d'\\
 c'-2 d' & 2d' 
\end{array}
\right)   
\left( 
\begin{array}{c} 
\mqs \\
\mqv
\end{array}
\right)   \, .
\label{massfit_generic}
\end{eqnarray}

We have performed semiquenched fits to eqs.\ref{quarkmatrix} and
\ref{massfit_generic} using the subset of mesonic data with
$\kappa_{\rm sea} = \kappa_{\rm val_1} = \kappa_{\rm val_2},
\kappa_{\rm sea} = \kappa_{\rm val_1} \neq \kappa_{\rm val_2}, 
\kappa_{\rm sea} \neq \kappa_{\rm val_1} = \kappa_{\rm val_2}$
, as described by
the $\kappa$ combinations in table \ref{kappas1}.  In order to ensure
consistency with the above light sector analysis we have used the
parameters  from the symmetric fits as inputs, namely
$a'=2b, \kappa_{sea}^{light},\kappa_{sea}^{c}, m'^{crit}=m^{crit},
c'=c$.

\begin{figure}
\begin{center}
\noindent\parbox{14cm}{
\parbox{7cm}{\epsfxsize=7cm\epsfbox{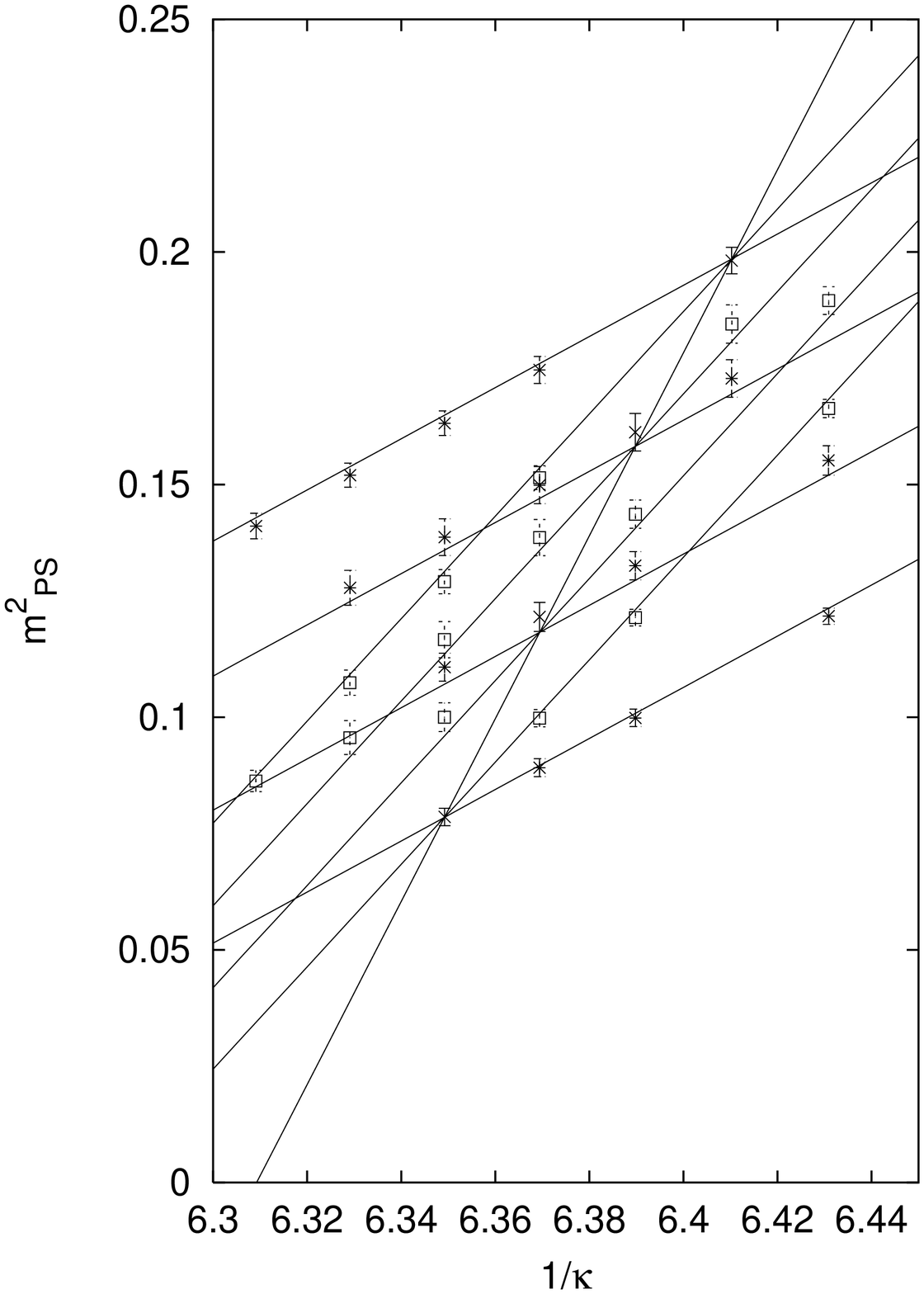}}
\nolinebreak\hfill
\parbox{7cm}{\epsfxsize=7cm\epsfbox{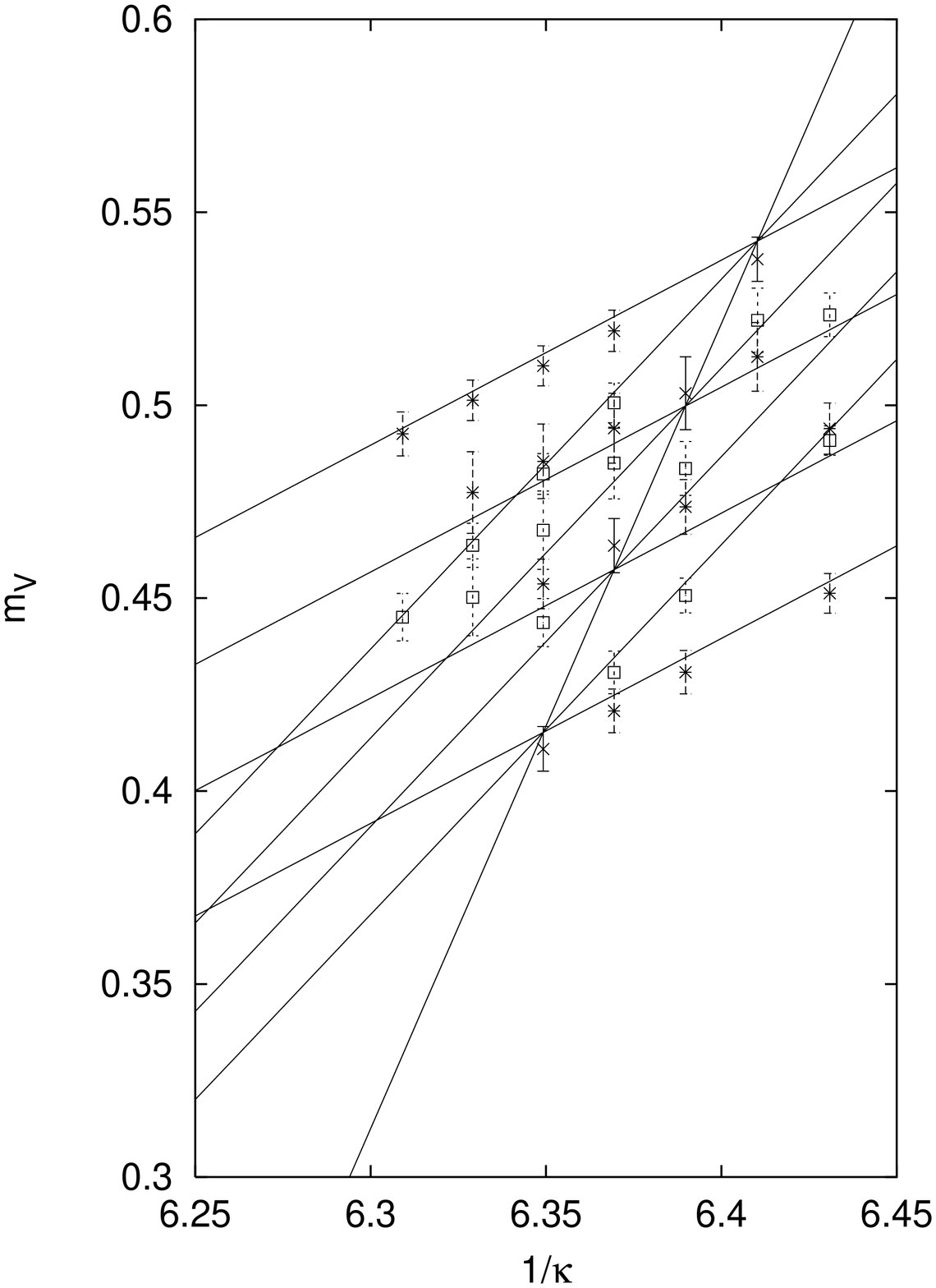}}}
\caption{Simultaneous fit of all pseudoscalar-data and vector-data to   
\eq\ref{quarkmatrix}. Symbols: $* = \mot$; $\Diamond = \moth$;
  $\Box = \mtf$.\label{fig_hugefit}}
\end{center}
\end{figure}

The results of such simultaneous fitting are listed in table
\ref{s_chiral} and illustrated in the plots of fig. \ref{fig_hugefit}.
All fits are characterized by  reasonable
$\chisq$. As we will discuss below, we have tested the  stability
of the procedure by relaxing the constraints.

It is obvious how to extend the analysis to nonlinear contributions in
$m_{sea}$ and $m_{val}$: 
\beq
m = m'^{\rm crit} + (c' - 2d')m_{\mqs} +
2 d'\mqv  + (e'-f'-g')\mqs^2
+ f'\mqs\mqv + g'\mqv^2
\label{thorsten}
\eeq The parameters from this nonlinear ansatz, with equal constraints
from the above symmetric analysis, and with $e'=e$, are also
included in table
\ref{s_chiral}. Note that the coefficients of $\mqs^2$ and $\mqs\mqv$
both come out to be negative, while the prefactor of $\mqv^2$ is positive
but small.

\begin{figure}
\begin{center}
\epsfxsize=14cm\epsfbox{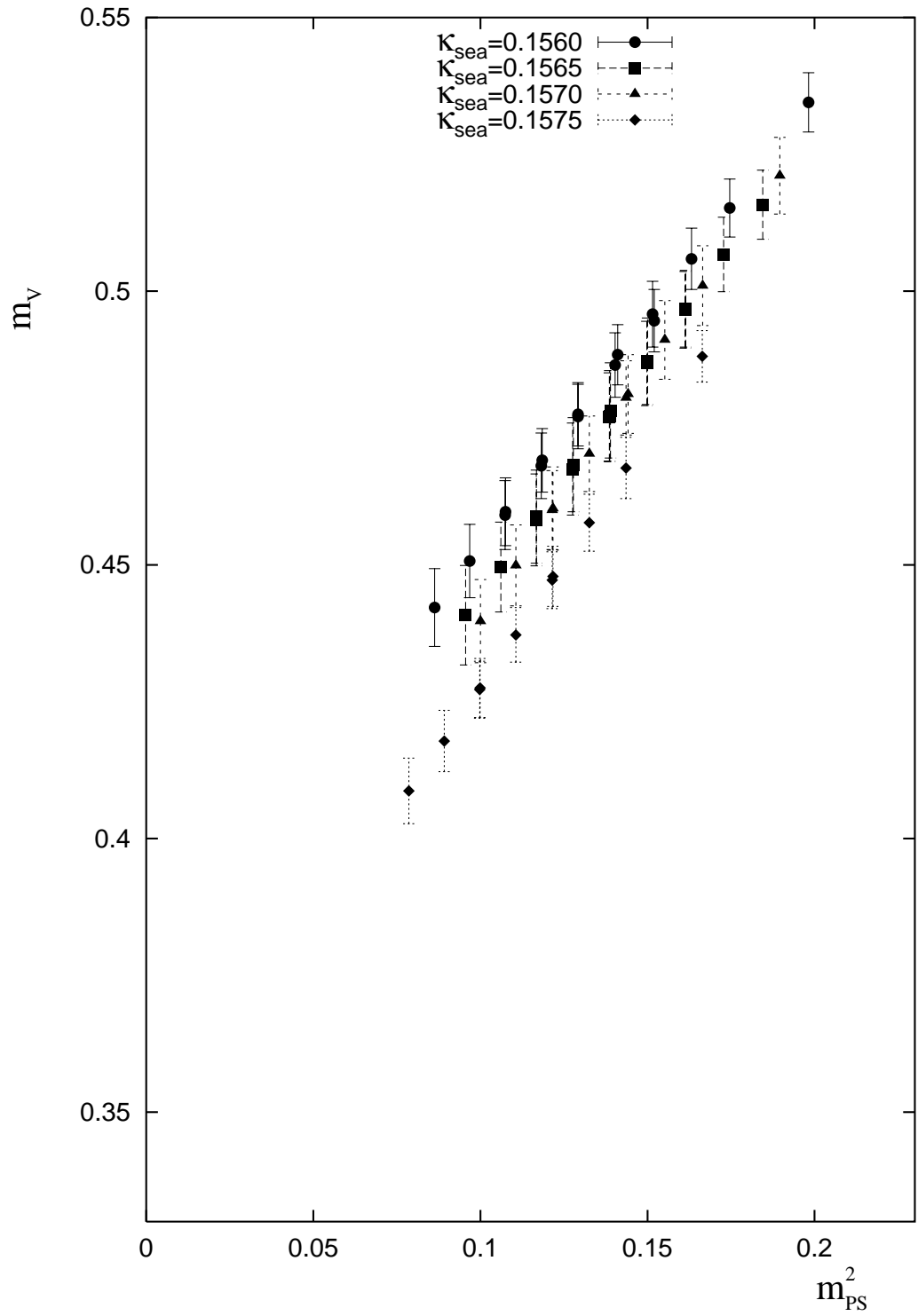}
\caption{Mass of the vector meson as a function 
of $m_\pi^2$.\label{fig_mv_vs_mps2}}
\end{center}
\end{figure}

Motivated by chiral perturbation theory and quenched QCD one might
expect, instead of eq.\ref{massfit_generic}, a direct connection on
the pseudoscalar mass, according  to the form
\begin{equation}
 m = m^{\rm crit} + \tilde{b}\,\mps \:\;,
\label{mv_mps_ansatz}
\end{equation}
which amounts  to restricting the parameters 
\begin{equation}
 c' = \tilde{b}\,a' \; , \; d'= \tilde{b}\,b' \; .
\label{massfit_constraints}
\end{equation}
To check for the validity of this idea, we have entered our entire
data set into a `scatterplot' with axes $m_V$ and $m_{PS}^2$.
Fig. \ref{fig_mv_vs_mps2} reveals that the entries do not collapse to
a single line but rather exhibit a clear pattern of sea quark mass
dependence, thus ruling out the one-slope ansatz
eq. \ref{mv_mps_ansatz}\footnote{ Obviously this statement can be
generalized to any dependence of type $m = f(m_{PS})$.}.
%
\subsubsection{Determination of $\kappa^{\rm strange}$ \label{strange}}

There are  three options to fix  $\kst$ from the spectrum:
\begin{itemize}
\item
from the $K^*$ mass  by solving
\beq
{m_{\rm V, {\sf sv}}(\ksl,\kst) \over m_{\rm V, {\sf ss}}(\ksl)} =
{M_{K^*} \over M_{\rho}} = 1.16 \; ,
\label{st1}
\eeq where $\kl$ is given by \eq\ref{klight}, 
\item 
or from the  Kaon mass  by matching
\beq
{m_{\rm PS, {\sf sv}}(\ksl,\kst) \over m_{\rm PS, {\sf ss}}(\ksl)} =
{M_{K} \over M_{\pi}} = 3.61 \; ,
\label{st2}
\eeq
\item
or  from the $\Phi$ meson mass according to
\beq
\label{st3}
{m_{V_{1}, {\rm vv}}(\ksl,\kst) \over  m_{V_{2}, {\rm ss}}(\ksl)} =
  {M_{\phi} \over M_{\rho}} = 1.326 \; .
\eeq
\end{itemize}

%


It is well known that quenched simulations with Wilson fermions fail
to reproduce the size of the experimental hyperfine splitting
among $K$ and $K^*$.
According to the results of the CP-PACS collaboration\cite{yoshie}, in
the continuum limit and on large lattices, the value of $M_{K^*}$
($M_\Phi$) turns out to deviate by 3\% (5\%) from experiment when
matching $\kappa^{strange}$ to $M_K$.  On the other hand they find
$M_{K^*}$ in accord with experiment when  using $M_\Phi$ as input
instead.  The deviation is generally attributed to quenching errors.

In the context of  the linear ansatz, the determination of
$\kappa^{strange}$ from these alternative scale choices proceeds
directly by explicit use of the fit parameters of table
\ref{s_chiral}.  Table \ref{strangefits} lists the resulting values.
While the two vector conditions, eqs. \ref{st1} and \ref{st3}, lead to
consistent results,  the $K$ meson mass condition asks for a
considerably  larger value of $\kst$. 

With the numbers for $\kst$ from this table one can proceed to compute
the meson masses in the strange sector, within the linear ansatz.  The
results are collected in table \ref{fixedseastrange}.  We find that,
contrary to the expectation, the discrepancy between the lattice
results and  the experimental hyperfine splitting remains
largely unaltered under unquenching.  One might be tempted to blame
the linear ansatz for this failure.  However, as can be seen from
table \ref{strangefits}, the spread in $\kst$ is by no means decreased
under a quadratic extrapolation.  We shall come back to this point
when we discuss the $J$-parameter.

If one interprets the spread from the three strange quark mass
settings as a systematic error, our `best' value for the strange hopping
parameter reads: 
\beq
\kst = 0.15608(14)(46)\; ,
\eeq
which agrees 
with the mass ratios
\beq {m_{K^*} \over m_K } = 1.59(17) \quad \quad
\mbox{and} \quad \quad {m_\Phi \over m_K} = 1.78(22) \; .
\eeq
The experimental mass ratios are $M_{K^*}/M_K=1.8$
and $M_\Phi/M_K=2.06$.
The quoted value of $\kst$ implies a strange quark 
\beq 
m^{\rm strange}_{\rm \overline{MS}}(2{\rm GeV}) = 151(30) {\rm MeV}
\eeq to be
compared to our previous value from the 3 sea quark
analysis\cite{qmasses} 
$m^{\rm strange}_{\rm \overline{MS}}(2{\rm GeV}) = 140(20) {\rm MeV}$.

The decay constants $f_{K}$ and $1/f_{\phi}$ can be determined
using the semi-quenched ansatz, eq.\ref{massfit_generic}, with
$m$ replaced by $f_{PS}$ and $1/f_V$ respectively. We compile the
results in table \ref{fixedseastrange}. It turns out that the
conditions, eqs. \ref{st1}, \ref{st2} and \ref{st3},  lead to
consistent answers, the spread of $3\%$ being well covered by the statistical
uncertainty.

\subsubsection{Stability of the semiquenched analysis}
By lifting the constraint one can convince oneself
in two ways of the stability of the light
sector, with respect to feedback from the strange sector:

{\it (i)}  
performing an unconstrained fit to eq.  \ref{quarkmatrix}, ond finds
(with $\chisq = 22/29$) for the critical value of $\kappa_{sea}$ \beq
\ksc = 0.158497 \err{47}{43} \, , \eeq which is nicely consistent with
the result from the symmetric analysis, eq.  \ref{eq_kcsym}.

{\it (ii)}  
An equally satisfying result is achieved with respect to
$\ksl$: the vector masses are reproduced with $\chisq = 17/29$, and one
obtains the value of the light hopping parameter as \beq \ksl =
0.158451 \err{47}{41} \, .
\label{eq_strl}
\eeq This number is also in good  agreement with the outcome of the
symmetric analysis, as given in eq. \ref{eq_klight}.  
\section{Discussion}
\subsection{The J-parameter}

In the quenched scenario the dimensionless parameter \beq J = M_{K^*}
{ d M_V \over d M^2_{PS}} \; \eeq has been proposed as a 
suitable lattice
observable to  avoid chiral extrapolations altogether \cite{Jpar}, on the
level of the single mass (i.e. effective $\kappa$) dependence of the
approximation.  Assuming the vector and pseudoscalar trajectories to
be linear one can estimate the slope within this expression from
experimental mass ratio, such that \beq J^{\sf exp} \simeq M_{K^*}
{M_{K^*} - M_{\rho} \over M_K^2 - M_\pi^2} = 0.48(2) \quad .
\label{eq_j_exp}
\eeq The quenched lattice value for $J$ generally is of the
order of $.38$, i.e. 25 \% below the empirical estimate.  It has
generally been surmised that this discrepancy provides evidence of a
quenching error (see e.g. the review \cite{yoshie}).

Our approach to $N_f = 2$ full QCD treats light and strange quarks on
unequal footing, as we associate them with sea and valence quarks,
respectively. Basically this induces a genuine two-parameter
dependence of $m_V$ on the light and strange quark masses.  Thus, in
the $N_f = 2$ theory, the notion of {\it one} effective hopping
parameter is not appropriate and the lattice determination of $J$ {\it
does not eo ipso enjoy } the merit of avoiding chiral extrapolation.

We have demonstrated in section \ref{strange} that our data does not
confirm the single slope ansatz, eq.\ref{mv_mps_ansatz}.  Thus, in
order to avoid the problem of choosing an effective slope $ d M_V
\over d M^2_{PS}$, we calculate the lattice value of $J$ directly from
the experimental definition, i.e. we insert our lattice masses,
c.f. tabs.  \ref{fixedseastrange},\ref{table_latt}, on the r.h.s. of
eq.\ref{eq_j_exp}. We find
\begin{equation}
J = 0.33(3)
\label{eq_J_st2}
\end{equation}  
if we define the strange quark mass by eq.\ref{st2}, and
\begin{equation}
J = 0.32(3)
\label{eq_J_st1}
\end{equation}
for the condition eq.\ref{st1}.
Both values are well below the quenched results.
 
We compare this result with the outcome from an analysis restricted to
the symmetric data. In this case an effective slope value $\tilde{b} = {d M_V
\over d M^2_{PS}}$ can be determined by a linear fit to
$m_V(m_{PS}^2)$ on the symmetric data set. The result is $\tilde{b}_{\sf sym}
= 1.07(6)$.  Following ref.\cite{Jpar} we set the quark mass by the
condition $m_V$ = 1.8$m_{PS}$, which corresponds to the experimental
$K^*/K$ mass ratio. This then produces the estimate $J$=0.40 (2),
which is significantly above the result of our two-slope analysis,
eqs. \ref{eq_J_st2} or \ref{eq_J_st1}.  We disfavour this
approach, however, since the (sea) quark mass, which fulfills the
condition $m_V$ = 1.8$m_{PS}$ on the symmetric line is purely
effective and does not correspond to a sea of light $u$ and $d$
quarks.

A third possible way to estimate  $J$ in the context of linear 
extrapolations is to apply
the above procedure not to the symmetric line, but on each individual line
of  {\it fixed sea quark mass}, with subsequent 
extrapolation of $J$ in $\kappa_{sea}$. In this approach, however, one
has to artificially impose the physical condition $m_V$ = 1.8$m_{PS}$ 
on  each one  of the unphysical sea quark values.
One can argue that a possible sea quark effect
could be easily washed out by such unnaturally
guided  procedure. 
And indeed: we do not recover any appreciable  dependence on the 
dynamical quark mass with this method. Moreover, the numbers, 
$J(\kappa_{sea}=0.1560)= 0.34(5)$, $J(\kappa_{sea}=0.1565)= 0.35(4)$,
$J(\kappa_{sea}=0.1570)= 0.35(5)$, $J(\kappa_{sea}=0.1575)= 0.36(5)$
come out close to the quenched values.

Needless to say, the $J$-analysis does not provide us with independent
information: obviously if we had succeeded in predicting the
experimental Kaon and $K^*$ masses with a single value of $\kst$, the
results for $J$, eqs. \ref{eq_J_st2},\ref{eq_J_st1}, would agree
exactly with $J^{\sf exp}$.  One might blame the linear ansatz for the
failure of $J$ and $J^{\sf exp}$ to coincide.  In this sense this
feature might be considered as {\it an evidence for curvature} in the
vector particle trajectory. In order to explore this possibility, we
have carrried out additional {\it quadratic fits} to the vector
particle trajectory both on the symmetric  and on the full data
set. The results for the fit parameters can be found in tables
\ref{table_pr} and \ref{s_chiral}. The coefficients of the quadratic
terms turn out to be negative, albeit {\it zero} within the
errors.  This then {\it even lowers} the  value of $J$!

As yet another alternative we have also used an ansatz with
the next to linear order in the quark mass $\sim m^{3/2}$.  Such a
behaviour is expected by chiral perturbation
theory\cite{chiral_pbt}. However, we again find that the $J$ parameter
decreases compared to the result of the linear ansatz.
\subsection{Consistency of scale determinations}
The lattice numbers for masses and decay constants can be translated
into physical results once the lattice cutoff $a^{-1}$ at $\beta=5.6,
N_F=2$ has been determined. This is done by matching the lattice
number of one observable with its experimental counterpart.
Obviously, within  a complete numerical solution of QCD, the size of
$a^{-1}$ should be independent of the choice of the particular
observable selected  to set the scale.  {\it Vice versa}, a variation of
the cutoff with the observable provides another measure for the
systematic uncertainty of our lattice calculation.

Table \ref{table_latt} exhibits the values of the cutoff as obtained
by matching physical scales inside the light sector: the $\rho$ mass,
the nucleon mass and the pion decay constant.  Within statistical
errors, all results appear consistent.  The difference between
$a^{-1}_N$ and $a^{-1}_{f_\pi}$ however reflects a systematic
uncertainty of $\simeq 20\%$, this being of course related to the
the error from the chiral extrapolation, c.f. section \ref{s_light}.

The impact of the uncertainty due to the chiral extrapolation can also
be demonstrated on $a^{-1}_\rho$ itself by making quadratic fits to
the vector trajectory (see table \ref{table_latt}).  One observes a 10
\% change in $a_\rho^{-1}$ which goes along with an amplification of
the error under the quadratic extrapolation from 4 to 15 \%.

The physical predictions are collected in table \ref{fixedsea1} for
the light sector, and in table \ref{fixedsea2} for the masses
of particles
containing strange quarks. Here we used $a_\rho^{-1}$ both from linear
and quadratic fits in order to test for the stability.  It turns out
that the admission of quadratic contributions in fitting the vector
trajectory increases both baryon masses by 10 to 15 \% and does not
reproduce the experimental $N-\Delta$ splitting.
The uncertainty in the strange sector is clearly dominated by
the mismatch of $\kappa^{\rm strange}(K)$ and $\kappa^{\rm
  strange}(K^*)$ which is connected with the failure to predict the
experimental $K-K^*$ splitting. 
The physical results for $f_K$ and $1/f_{\phi}$ are listed in table
\ref{fixedsea3}. As we mentioned above, the uncertainty due to
the choice of  $\kappa^{\rm strange}$ is covered by the statistical
errors in this case.

The analogous quenched results 
($\beta_q = 6.0$, $a_\rho^{-1}= 2.3\mbox[GeV]$, 200 configurations of
$16^3\times 32$ lattices) are contained in table \ref{quenched}.
Notice that the errors on the nucleon and $\Delta$ masses are
smaller by a factor 2 to 4.

\begin{figure}
\begin{center}
\epsfxsize=13cm\epsfbox{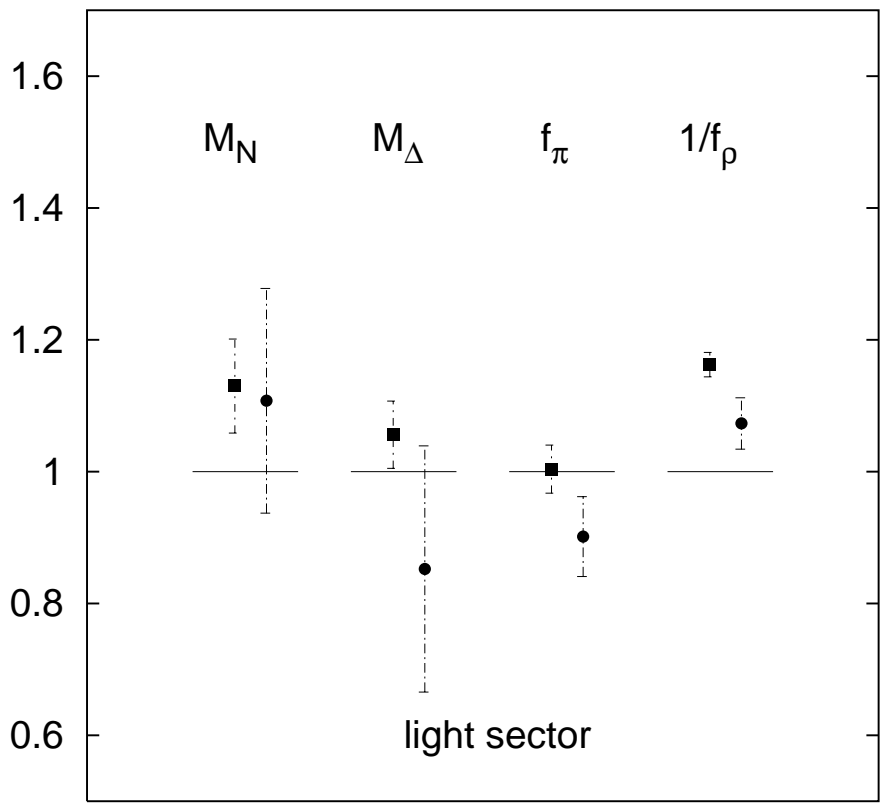}
\caption{Comparison of $N_F=2$ (circles) and quenched results
(squares) in the light sector with experiment.
The data is normalized to its experimental
values, namely $M_N = 938$MeV, $M_\Delta = 1232$MeV, $f_\pi=132$MeV,
$1/f_\rho=0.199\sqrt{2}$. To set the scale we used the linear fit to
the vector meson trajectory. 
\label{fig_compare_full_quen_light}}
\end{center}
\end{figure}

\begin{figure}
\begin{center}
\epsfxsize=13cm\epsfbox{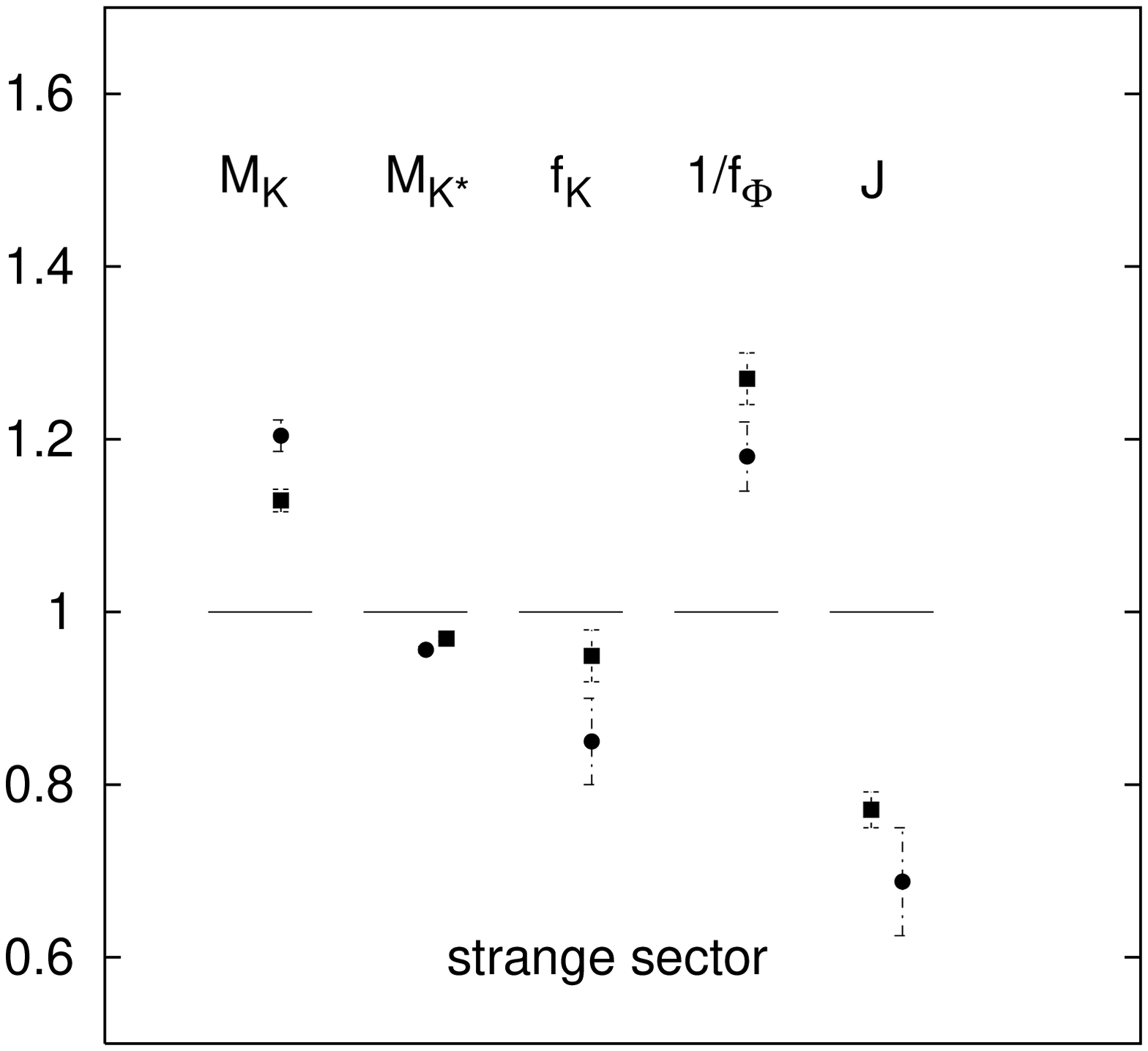}
\caption{Comparison of $N_F=2$ (circles) and quenched results
(squares) in the strange sector with experiment.
The data is normalized to its experimental
values, namely $M_K = 495$MeV, $M_{K^*} = 892$MeV, $f_K=160$MeV,
$1/f_\Phi=0.234$ and $J=0.48$.
To set the scale we used the linear fit to
the vector meson trajectory. 
\label{fig_compare_full_quen_strange}}
\end{center}
\end{figure}

In figs. \ref{fig_compare_full_quen_light} and
\ref{fig_compare_full_quen_strange} we present a compilation of the
various quantities. We conclude that the data resists to reveal clear
sea quark effect on these observables. In particular there remains the
problem to account for the $N-\Delta$ mass splitting.  It is unlikely
that an increase in statistics would remedy the situation.

\section{Summary and Conclusions}
We have presented, on moderately sized lattices and at fixed $\beta$,
a detailed analysis of the light and strange hadron spectra in full
QCD.  Both meet their particular difficulties: while the strange
spectrum calculation is hampered by the technical requirement of $N_F
= 2$, the light baryonic sector faces  the problem   of
considerable variation of hadron masses under chiral extrapolation.

We found that with these limitations we are not able to overcome the
well-known shortcomings of quenched calculations, namely the
underestimation of the $K-K^*$ and $N-\Delta$ spin splittings.

The experimental $J$-parameter cannot be explained in a linear
scenario of vector trajectories and the admission of higher order
terms does not help to improve on the situation.  In view of this
result, it would be highly desirable to make more realistic
computations by  including a {\it third} type of active sea quark.

The issue  of the $N-\Delta$ splitting could be considerably clarified
by closer approach to the chiral limit, on larger lattices. Work along
these lines is in progress\cite{tchl}.\\

{\bf Acknowledgements}\\
This work was done with support from the DFG grants Schi 257/1-4,
 257/3-2, 257/3-3 and the DFG Graduiertenkolleg ``Feldtheoretische und
 Numerische Methoden in der Statistischen und
 Elementarteilchenphysik''.  G.R.  gratefully acknowledges the
 Max-Kade Foundation for support. The HMC productions were run on
 the APE100 hardware at IfH Zeuthen and the  Quadrics machine provided
by DFG to the Schwerpunkt ``Dynamische Fermionen'', operated by 
 the University of Bielefeld. Most of the hadron analysis was done on
 the CRAY T3E systems of ZAM at FZJ.  We thank the support teams of
 these institutions.  We thank G. Siegert for her contributions during
 the early stages of this work.

Last but not least, we are most grateful to the members of the INFN
 groups in at the Universities of Rome and Pisa, in particular
 R. Tripiccione and F. Rapuano for friendly cooperation; without the
 commitment of the APE team this work would not have been possible.


%
%
%
%
\newpage
\clearpage
%
\appendix{}
\section{Renormalization constants\label{app_renorm}}
We briefly present our method of choice for the extraction of the
renormalization constants $Z_{A,V}$. We use the tadpole improved
perturbation theory results from Lepage and Mackenzie\cite{lepage}.
 The procedure 
is as follows: 
\begin{itemize}
\item
Use the plaquette values (see table \ref{alpha_value})
to calculate the value of
$\alpha_V({3.41 \over a})$ using:
\begin{equation}
-\log \langle \frac{1}{3} {\rm Sp} \Box \rangle=
\frac{4 \pi}{3} \alpha_V(3.41/a) ( 1 - (1.191+0.025 n_f) \alpha_V )  .
\end{equation}
\item
Convert to $\overline {MS}$ scheme using
\begin{equation}
\frac{1}{\alpha_{\overline{MS}}(3.41/a)} 
= \frac{1}{\alpha_V(3.41/a)} + 0.822  .
\end{equation}
\item
Run to a scale ${1 \over a}$.
\item
Use tadpole improved perturbation theory:
\beq
Z_A = 1 - 0.31 \alpha_{\overline{MS}}({ 1 \over a})\, , \hspace{1cm} 
Z_V = 1 - 0.82 \alpha_{\overline{MS}}({ 1 \over a}) 
\eeq
\item
Neglecting the light quark dependence of the plaquette we find
\begin{equation}
Z_A = 0.93 \quad, \quad Z_V = 0.82 \;
\end{equation}
for $n_f=2$, and $Z_A=0.94,Z_V=0.83$ in the quenched case.
We also need to rescale our quark fields:
\begin{equation}
\sqrt{2\kappa}\Psi \to  \sqrt{1-3 \kappa/4\kappa_c} \Psi .
\end{equation}
Matrix elements in the appendix are listed without the rescaling of
the quark fields and before applying the renormalization constants. 
\end{itemize}
%
%
%
%
%
\begin{table}[t]
\begin{center}
\caption{Simulation parameters and characteristic numbers.
\label{simulation}} 
\hspace{0.2cm}
\begin{tabular}{|ccccccc|}
\multicolumn{7}{|c|} {$\beta  = 5.6$, $N_f = 2$, $V \times T
  = 16^3 \times   32$}\\
\hline
$\kappa_{\rm sea}$ & 0.156 &  0.1565 & \multicolumn{2}{c}{0.1570} &
  \multicolumn{2}{c|}{0.1575} \\ 
Algorithm    &  o/e     & SSOR  & o/e       & SSOR        &  o/e & SSOR\\ 
\hline
$T$                & 1     & 1      & 1      & 1 & 1 & 0.5\\
$N_{md}\pm\sigma(N_{md})$  & $100\pm 20$ & $100\pm 20$ &  $100\pm 20$
  & $100\pm 20$ & $100\pm 20$ & $71\pm 12$ \\
$N_{CSG}$          & 6 & 7 & 8 & 9 & 11 & 3 \\
\# of iter.        & 85(3) & 89(6) & 168(5) & 125(3) & 317(12)& 150(6)\\
acc. rate[\%]      & 85    & & 84     & 80     & 76     & 73\\
\# of traj.        & 5000  & 5000 & 1500   & 3500   & 3000  & 2000 \\   
\hline
\# of confs.  & 200 & 200 & \multicolumn{2}{c}{200} &
  \multicolumn{2}{c|}{200} \\   
\hline
$\kv - \kv$ comb. & 15 &15 &\multicolumn{2}{c}{10} &\multicolumn{2}{c|}{10} \\
\end{tabular}
\end{center}
\end{table}
%
%
%
\begin{table}
\begin{center}
\caption{The operators studied.\label{ops}}
\hspace{0.2cm}
\begin{tabular}{|ll|}
Mesons        &   $\chi_A^\dagger(x)\chi_A(0)$ \\ \hline
Pseudoscalar:  &   $\chi_{PS}(x) = P_5 = \bar{q}'(x) \gamma^5 q(x)$\\
Vector:        &   $\chi^{\mu}_{V}(x) = V^\mu = \bar{q}'(x) \gamma^{\mu}
q(x)$\\
Scalar:        &   $\chi_{Sc}(x) = \bar{q}'(x) q(x)$\\
Axial-vector:   &   $\chi_{Ax}(x) = A^\mu = \bar{q}'(x) \gamma_5 \gamma^{\mu}
q(x)$\\
\hline \hline
Baryons       &   $\chi_A^\dagger(x)\chi_A(0)$ \\ \hline
Nucleon:       & $\chi_{N}(x) = \epsilon _{abc} ( q_a C \gamma_5 q_b)
q_c$\\
$\Delta$:      & $\chi^{\mu}_{\Delta}(x) = \epsilon _{abc} ( q_a  
C \gamma^{\mu} q_b) q_c$\\ 
\hline \hline
Decay Constants       &   $\chi_A^\dagger(x),\chi_B(0)$ \\ \hline
Pseudoscalar:  &   $(A_0^\dagger,P_5), (P_5^\dagger,A_0), (A_0^\dagger,A_0)$  \\
Vector:        &   $(V_i^\dagger, V_i)$ \\
\end{tabular}
\end{center}
\end{table}
%
%
\begin{table}
\begin{center}
\caption{The run parameters for $\kappa_{val}$ \label{kappas1}}
\hspace{0.2cm}
\begin{tabular}{|c|c|}
$\ks$   & $ \{ \kv \}$ \\ \hline
0.156 & $\{0.156, 0.157, 0.1575, 0.158, 0.1585 \}$ \\ \hline
0.1565 & $\{0.156, 0.1565, 0.157, 0.1575, 0.158 \}$ \\ \hline
0.157 & $\{ 0.1555, 0,1565, 0.157, 0.1575 \}$ \\ \hline
0.1575 & $\{ 0.1555, 0,1565, 0.157, 0.1575 \}$ \\ 
\end{tabular}
\end{center}
\end{table}
%
%
\begin{table}
\begin{center}
\caption{Integrated autocorrelation times $\tau_{\rm{int}}$ for
  pseudoscalar, vector and nucleon for smeared-local and
  smeared-smeared correlators (numbers are in units of HMC time).
\label{auto}}
\hspace{0.2cm}
\begin{tabular}{|ccc|cc|cc|cc|c|}
$\beta$ & $\kappa$ & $V$ & \multicolumn{2}{c}{$\tau_{int}(M_{PS})$} 
& \multicolumn{2}{c}{$\tau_{int}(M_V)$} 
& \multicolumn{2}{c}{$\tau_{int}(M_N)$} & $B$ \\ \hline
        &      &    & sl & ss & sl & ss & sl & ss & \\ \hline \hline
5.6 & 0.1560   & $16^3 \times 32$ & 
22(9) & $<25$  & $<25$ & $<25$ & $<25$ & $<25$ & 6 \\ \hline 
5.6 & 0.1570   & $16^3 \times 32$ & 
19(6) & 17(5) & $<25$ & $<25$ & $<25$ & $<25$ & 6 \\ \hline
5.6 & 0.1575   & $16^3 \times 32$ & 
44(20) & 33(22) & $<25$ & $<25$ & 37(20) & 32(24) & 7 \\ 
\end{tabular}
\end{center}
\end{table}
%
%


\begin{table}
\begin{center}
\caption{\label{raw_mass_156}Lattice results for the masses of
Pion, Rho, Nucleon and Delta at $\kappa_{\rm sea} = 0.156$. For all
fits we find $0.4 \leq \chi^2/\mbox{d.o.f} \leq 1$.}
\begin{tabular}{|ccccc|}
\hline 
\hline
\multicolumn{5}{|c|}{$\kappa_{\rm sea} = 0.156$}\\\hline
\multicolumn{5}{|c|}{nconfigs $ = 198$, nboot $= 200$, correlated}\\\hline
$(t_{\rm min}$,$t_{\rm max})$ & (9,15) & (9,15) & (9,14) & (8,14) \\
\hline 
\hline
$\kappa_1$-$\kappa_2$ & $m_{\pi}$ & $m_{\rho}$ & $m_{N}$ & $m_{\Delta}$ \\
\hline
 0.1585-0.1585 & $0.2937\errr{ 41}{ 42}$ & $0.4422\errr{ 64}{ 71}$ & $0.676\errr{ 19}{ 16}$ & $0.788\errr{ 21}{ 19}$ \\
\hline
 0.1585-0.1580 & $0.3111\errr{ 43}{ 41}$ & $0.4507\errr{ 62}{ 67}$ & $0.689\errr{ 17}{ 15}$ & $0.796\errr{ 18}{ 18}$ \\
\hline 
 0.1580-0.1580 & $0.3277\errr{ 41}{ 40}$ & $0.4591\errr{ 58}{ 63}$ & $0.711\errr{ 16}{ 13}$ & $0.809\errr{ 17}{ 17}$ \\
\hline
 0.1585-0.1575 & $0.3279\errr{ 41}{ 39}$ & $0.4597\errr{ 58}{ 62}$ & $0.702\errr{ 15}{ 14}$ & $0.804\errr{ 18}{ 17}$ \\
\hline
 0.1580-0.1575 & $0.3438\errr{ 39}{ 38}$ & $0.4681\errr{ 60}{ 56}$ & $0.724\errr{ 15}{ 13}$ & $0.817\errr{ 16}{ 16}$ \\
\hline
 0.1585-0.1570 & $0.3443\errr{ 39}{ 38}$ & $0.4691\errr{ 58}{ 57}$ & $0.715\errr{ 15}{ 13}$ & $0.812\errr{ 16}{ 17}$ \\
\hline
 0.1575-0.1575 & $0.3594\errr{ 36}{ 36}$ & $0.4771\errr{ 59}{ 55}$ & $0.747\errr{ 14}{ 12}$ & $0.832\errr{ 16}{ 17}$ \\
\hline
 0.1580-0.1570 & $0.3595\errr{ 36}{ 36}$ & $0.4775\errr{ 58}{ 55}$ & $0.737\errr{ 14}{ 12}$ & $0.825\errr{ 16}{ 16}$ \\
\hline
 0.1575-0.1570 & $0.3745\errr{ 34}{ 34}$ & $0.4865\errr{ 59}{ 54}$ & $0.760\errr{ 13}{ 12}$ & $0.840\errr{ 15}{ 16}$ \\
\hline
 0.1585-0.1560 & $0.3756\errr{ 36}{ 35}$ & $0.4884\errr{ 55}{ 53}$ & $0.742\errr{ 13}{ 13}$ & $0.830\errr{ 16}{ 16}$ \\
\hline
 0.1570-0.1570 & $0.3892\errr{ 33}{ 33}$ & $0.4958\errr{ 60}{ 51}$ & $0.783\errr{ 13}{ 11}$ & $0.857\errr{ 14}{ 16}$ \\
\hline
 0.1580-0.1560 & $0.3899\errr{ 33}{ 34}$ & $0.4969\errr{ 57}{ 51}$ & $0.764\errr{ 13}{ 12}$ & $0.843\errr{ 15}{ 15}$ \\
\hline
 0.1575-0.1560 & $0.4040\errr{ 33}{ 31}$ & $0.5059\errr{ 56}{ 55}$ & $0.786\errr{ 12}{ 12}$ & $0.858\errr{ 13}{ 16}$ \\
\hline
 0.1570-0.1560 & $0.4179\errr{ 35}{ 30}$ & $0.5152\errr{ 51}{ 53}$ & $0.808\errr{ 11}{ 11}$ & $0.875\errr{ 14}{ 15}$ \\
\hline
 0.1560-0.1560 & $0.4452\errr{ 32}{ 29}$ & $0.5345\errr{ 54}{ 49}$ & $0.852\errr{10}{ 11}$  & $0.910\errr{ 12}{ 14}$ \\
\hline
\hline
\end{tabular}
\end{center}
\end{table}


\begin{table}
\begin{center}
\caption{\label{raw_mass_1565}Lattice results for the masses of
Pion, Rho, Nucleon and Delta at $\kappa_{\rm sea} = 0.1565$.
We find $1.5 \leq \chi^2/\mbox{d.o.f}\leq 3$ for fits to $\pi$ and $\rho$ 
, and $0.5 \leq \chi^2/\mbox{d.o.f}\leq 1.5$ for fits to nucleon
 and $\Delta$.}
\begin{tabular}{|ccccc|}
\hline 
\hline
\multicolumn{5}{|c|}{$\kappa_{\rm sea} = 0.1565$}\\\hline
\multicolumn{5}{|c|}{nconfigs $ = 198$, nboot $= 200$, correlated}\\\hline
$(t_{\rm min}$,$t_{\rm max})$ & (9,15) & (9,15) & (9,14) & (8,14) \\
\hline 
\hline
$\kappa_1$-$\kappa_2$ & $m_{\pi}$ & $m_{\rho}$ & $m_{N}$ & $m_{\Delta}$ \\
\hline
 0.1580-0.1580 & $0.3092\errr{ 61}{ 50}$ & $0.4408\errr{ 91}{ 89}$ &
 $0.656\errr{ 19}{ 21}$ & $0.704\errr{ 37}{ 36}$ \\
\hline
 0.1580-0.1575 & $0.3257\errr{ 59}{ 51}$ & $0.4496\errr{ 82}{ 81}$ &
 $0.673\errr{ 20}{ 20}$ & $0.721\errr{ 35}{ 34}$ \\
\hline 
 0.1575-0.1575 & $0.3416\errr{ 58}{ 51}$ & $0.4582\errr{ 84}{ 81}$ &
 $0.701\errr{ 18}{ 18}$ & $0.748\errr{ 33}{ 31}$ \\
\hline
 0.1580-0.1570 & $0.3418\errr{ 58}{ 52}$ & $0.4588\errr{ 85}{ 82}$ &
 $0.688\errr{ 19}{ 20}$ & $0.736\errr{ 34}{ 31}$ \\
\hline
 0.1575-0.1570 & $0.3572\errr{ 54}{ 53}$ & $0.4675\errr{ 84}{ 80}$ &
 $0.716\errr{ 17}{ 18}$ & $0.763\errr{ 31}{ 26}$ \\
\hline
 0.1580-0.1565 & $0.3575\errr{ 54}{ 52}$ & $0.4683\errr{ 86}{ 84}$ &
 $0.702\errr{ 18}{ 20}$ & $0.750\errr{ 34}{ 30}$ \\
\hline
 0.1570-0.1570 & $0.3723\errr{ 54}{ 52}$ & $0.4770\errr{ 81}{ 77}$ &
 $0.741\errr{ 17}{ 17}$ & $0.789\errr{ 26}{ 23}$  \\
\hline
 0.1575-0.1565 & $0.3724\errr{ 54}{ 53}$ & $0.4772\errr{ 83}{ 79}$ &
 $0.730\errr{ 17}{ 18}$ & $0.776\errr{ 28}{ 25}$ \\
\hline
 0.1570-0.1565 & $0.3871\errr{ 53}{ 51}$ & $0.4868\errr{ 77}{ 74}$ &
 $0.755\errr{ 17}{ 16}$ & $0.802\errr{ 26}{ 23}$ \\
\hline
 0.1580-0.1560 & $0.3728\errr{ 54}{ 52}$ & $0.4782\errr{ 87}{ 82}$ &
 $0.716\errr{ 18}{ 20}$ & $0.763\errr{ 32}{ 27}$ \\
\hline
 0.1565-0.1565 & $0.4016\errr{ 52}{ 50}$ & $0.4966\errr{ 70}{ 69}$ &
 $0.778\errr{ 16}{ 15}$ & $0.827\errr{ 23}{ 22}$  \\
\hline
 0.1575-0.1560 & $0.3873\errr{ 54}{ 51}$ & $0.4872\errr{ 79}{ 76}$ &
 $0.744\errr{ 18}{ 18}$ & $0.789\errr{ 26}{ 24}$ \\
\hline
 0.1570-0.1560 & $0.4017\errr{ 52}{ 50}$ & $0.4968\errr{ 70}{ 67}$ &
 $0.768\errr{ 17}{ 17}$ & $0.815\errr{ 24}{ 23}$ \\
\hline
 0.1565-0.1560 & $0.4157\errr{ 51}{ 46}$ & $0.5067\errr{ 68}{ 66}$ &
 $0.792\errr{ 16}{ 15}$ & $0.839\errr{ 23}{ 21}$ \\
\hline
 0.1560-0.1560 & $0.4295\errr{ 50}{ 42}$ & $0.5168\errr{ 63}{ 63}$ &
 $0.814\errr{ 15}{ 15}$ & $0.862\errr{ 20}{ 20}$ \\
\hline
\hline
\end{tabular}
\end{center}
\end{table}


\begin{table}
\begin{center}
\caption{\label{raw_mass_157}Lattice results for the masses of
Pion, Rho, Nucleon and Delta at $\kappa_{\rm sea} = 0.157$
We find $1.5 \leq \chi^2/\mbox{d.o.f}\leq 3$ for fits to $\rho$ and 
$\Delta$ 
, and $0.5 \leq \chi^2/\mbox{d.o.f}\leq 1.5$ for fits to $\pi$ and nucleon.}
\begin{tabular}{|ccccc|}
\hline 
\hline
\multicolumn{5}{|c|}{$\kappa_{\rm sea} = 0.157$}\\\hline
\multicolumn{5}{|c|}{nconfigs $ = 198$, nboot $= 200$, correlated}\\\hline
$(t_{\rm min}$,$t_{\rm max})$ & (8,15) & (8,15) & (9,14) & (9,14) \\
\hline 
\hline
$\kappa_1$-$\kappa_2$ & $m_{\pi}$ & $m_{\rho}$ & $m_{N}$ & $m_{\Delta}$ \\
\hline
 0.1575-0.1575 & $0.3163\errr{ 53}{ 45}$ & $0.4397\errr{ 72}{ 76}$ &
 $0.695\errr{ 15}{ 19}$ & $0.774\errr{ 13}{ 14}$ \\ 
\hline
 0.1575-0.1570 & $0.3328\errr{ 50}{ 42}$ & $0.4499\errr{ 73}{ 74}$ &
 $0.708\errr{ 15}{ 15}$ & $0.786\errr{ 12}{ 14}$ \\ 
\hline 
 0.1570-0.1570 & $0.3486\errr{ 49}{ 41}$ & $0.4600\errr{ 70}{ 72}$ &
 $0.732\errr{ 14}{ 14}$ & $0.805\errr{ 13}{ 13}$ \\ 
\hline
 0.1575-0.1565 & $0.3489\errr{ 51}{ 40}$ & $0.4603\errr{ 72}{ 76}$ &
 $0.722\errr{ 15}{ 15}$ & $0.798\errr{ 12}{ 14}$ \\ 
\hline
 0.1570-0.1565 & $0.3641\errr{ 47}{ 37}$ & $0.4703\errr{ 67}{ 69}$ &
 $0.746\errr{ 14}{ 13}$ & $0.816\errr{ 14}{ 13}$ \\ 
\hline
 0.1565-0.1565 & $0.3790\errr{ 46}{ 38}$ & $0.4805\errr{ 64}{ 68}$ &
 $0.769\errr{ 12}{ 12}$ & $0.835\errr{ 14}{ 13}$ \\ 
\hline
 0.1575-0.1555 & $0.3797\errr{ 48}{ 37}$ & $0.4812\errr{ 66}{ 72}$ &
 $0.750\errr{ 14}{ 13}$ & $0.819\errr{ 14}{ 14}$ \\ 
\hline
 0.1570-0.1555 & $0.3939\errr{ 46}{ 36}$ & $0.4911\errr{ 61}{ 72}$ &
 $0.774\errr{ 13}{ 12}$ & $0.837\errr{ 14}{ 13}$ \\ 
\hline
 0.1565-0.1555 & $0.4080\errr{ 45}{ 34}$ & $0.5010\errr{ 63}{ 73}$ &
 $0.797\errr{ 12}{ 11}$ & $0.856\errr{ 13}{ 13}$ \\ 
\hline
 0.1555-0.1555 &$0.4354\errr{ 42}{ 30}$  & $0.5211\errr{ 61}{ 70}$ &
 $0.842\errr{10}{ 10}$ & $0.895\errr{ 14}{ 13}$ \\ 
\hline
\hline
\end{tabular}
\end{center}
\end{table}


\begin{table}
\begin{center}
\caption{\label{raw_mass_1575}Lattice results for the masses of
Pion, Rho, Nucleon and Delta at $\kappa_{\rm sea} = 0.1575$.
For all fits we find $0.5 \leq \chi^2/\mbox{d.o.f} \leq 1.5$.}
\begin{tabular}{|ccccc|}
\hline 
\hline
\multicolumn{5}{|c|}{$\kappa_{\rm sea} = 0.1575$}\\\hline
\multicolumn{5}{|c|}{nconfigs $ = 198$, nboot $= 200$, correlated}\\\hline
$(t_{\rm min}$,$t_{\rm max})$ & (7,15) & (10,15) & (8,14) & (8,14) \\
\hline 
\hline
$\kappa_1$-$\kappa_2$ & $m_{\pi}$ & $m_{\rho}$ & $m_{N}$ & $m_{\Delta}$ \\
\hline
 0.1575-0.1575 & $0.2803\errr{ 45}{ 23}$ & $0.4087\errr{ 55}{ 60}$ &
 $0.633\errr{12}{12}$ & $0.695\errr{ 16}{ 16}$ \\ 
\hline
 0.1575-0.1570 & $0.2986\errr{ 42}{ 20}$ & $0.4178\errr{ 56}{ 50}$ &
 $0.648\errr{10}{11}$ & $0.707\errr{ 15}{ 16}$ \\ 
\hline 
 0.1570-0.1570 & $0.3159\errr{ 37}{ 20}$ & $0.4272\errr{ 52}{ 50}$ &
 $0.673\errr{8}{9}$ & $0.730\errr{ 14}{ 14}$ \\ 
\hline
 0.1575-0.1565 & $0.3160\errr{ 37}{ 19}$ & $0.4275\errr{ 54}{ 51}$ &
 $0.662\errr{9}{ 10}$ & $0.719\errr{ 15}{ 15}$ \\ 
\hline
 0.1570-0.1565 & $0.3325\errr{ 33}{ 18}$ & $0.4372\errr{ 50}{ 51}$ &
 $0.687\errr{8}{9}$ & $0.742\errr{ 13}{ 14}$ \\ 
\hline
 0.1565-0.1565 & $0.3485\errr{ 30}{ 20}$ & $0.4472\errr{ 52}{ 47}$ &
 $0.712\errr{6}{7}$ & $0.764\errr{ 12}{ 13}$ \\ 
\hline
 0.1575-0.1555 & $0.3489\errr{ 31}{ 19}$ & $0.4479\errr{ 55}{ 48}$ &
 $0.690\errr{8}{9}$ & $0.743\errr{ 14}{ 15}$ \\ 
\hline
 0.1570-0.1555 & $0.3642\errr{ 30}{ 20}$ & $0.4577\errr{ 52}{ 48}$ &
 $0.715\errr{7}{8}$ & $0.765\errr{ 13}{ 13}$ \\ 
\hline
 0.1565-0.1555 & $0.3790\errr{ 28}{ 22}$ & $0.4677\errr{ 56}{ 47}$ &
 $0.738\errr{7}{7}$ & $0.786\errr{ 11}{ 13}$ \\ 
\hline
 0.1555-0.1555 & $0.4079\errr{ 27}{ 23}$ & $0.4881\errr{ 47}{ 43}$ &
 $0.785\errr{6}{7}$ & $0.828\errr{10}{ 12}$ \\ 
\hline
\hline
\end{tabular}
\end{center}
\end{table}

\begin{table}
\begin{center}
\caption{\label{raw_decay_156}Lattice results pseudoscalar and vector
meson decay constants at $\kappa_{\rm sea} = 0.156$. For all
fits we find $1 \leq \chi^2/\mbox{d.o.f} \leq 2$.}
\begin{tabular}{|ccccc|}
\hline 
\hline
\multicolumn{5}{|c|}{$\kappa_{\rm sea} = 0.156$}\\\hline
\multicolumn{5}{|c|}{nconfigs $ = 198$, nboot $= 200$, correlated}\\\hline
$(t_{\rm min}$,$t_{\rm max})$ & (9,15) &  & (9,15) &  \\
\hline 
\hline
$\kappa_1$-$\kappa_2$ & $\langle 0|A_0^l|\pi \rangle$ & $f_{\pi}/Z_A$
 & $ 3^{-1/2}\langle 0|V^l| V\rangle$ & $1/(Z_V f_{\rho})$ \\
\\
\hline
 0.1585-0.1585 & $0.0289\errr{ 11}{ 10}$  & $0.0985\errr{ 25}{ 26}$  &
 $0.1611\errr{ 82}{ 80}$  & $0.4758\errr{99}{99}$ \\ 
\hline
 0.1585-0.1580 & $0.0314\errr{ 12}{ 10}$  & $0.1008\errr{ 25}{ 27}$  &
 $0.1647\errr{ 76}{ 78}$  & $0.4680\errr{ 96}{99}$ \\  
\hline 
 0.1580-0.1580 & $0.0338\errr{ 12}{ 10}$  & $0.1030\errr{ 27}{ 29}$  &
 $0.1683\errr{ 67}{ 77}$  & $0.4609\errr{ 93}{99}$ \\  
\hline
 0.1585-0.1575 & $0.0337\errr{ 12}{ 10}$  & $0.1029\errr{ 27}{ 28}$  &
 $0.1685\errr{ 68}{ 77}$  & $0.4603\errr{ 94}{1e+02}$ \\  
\hline
 0.1580-0.1575 & $0.0362\errr{ 13}{ 11}$  & $0.1051\errr{ 28}{ 30}$  &
 $0.1722\errr{ 73}{ 75}$  & $0.4537\errr{ 94}{ 92}$ \\  
\hline
 0.1585-0.1570 & $0.0361\errr{ 12}{ 11}$  & $0.1049\errr{ 28}{ 28}$  &
 $0.1725\errr{ 72}{ 74}$  & $0.4526\errr{ 96}{ 95}$ \\  
\hline
 0.1575-0.1575 & $0.0386\errr{ 13}{ 12}$  & $0.1073\errr{ 28}{ 31}$  &
 $0.1763\errr{ 69}{ 75}$  & $0.4472\errr{ 90}{ 94}$ \\  
\hline
 0.1580-0.1570 & $0.0385\errr{ 13}{ 12}$  & $0.1072\errr{ 28}{ 30}$  &
 $0.1763\errr{ 68}{ 74}$  & $0.4465\errr{ 91}{ 95}$ \\  
\hline
 0.1575-0.1570 & $0.0410\errr{ 14}{ 12}$  &  $0.1094\errr{ 29}{ 31}$ &
 $0.1806\errr{ 66}{ 71}$  & $0.4405\errr{ 90}{ 92}$ \\  
\hline
 0.1585-0.1560 & $0.0408\errr{ 13}{ 12}$  & $0.1087\errr{ 27}{ 29}$  &
 $0.1804\errr{ 71}{ 73}$  & $0.4367\errr{ 96}{ 86}$ \\  
\hline
 0.1570-0.1570 & $0.0434\errr{ 14}{ 14}$  & $0.1116\errr{ 30}{ 32}$  &
 $0.1849\errr{ 64}{ 69}$  & $0.4343\errr{ 92}{ 88}$ \\  
\hline
 0.1580-0.1560 & $0.0433\errr{ 14}{ 13}$  & $0.1110\errr{ 29}{ 31}$  &
 $0.1846\errr{ 63}{ 71}$  & $0.4317\errr{ 92}{ 88}$ \\  
\hline
 0.1575-0.1560 & $0.0458\errr{ 14}{ 13}$  & $0.1133\errr{ 29}{ 32}$  &
 $0.1891\errr{ 63}{ 66}$  & $0.4266\errr{ 90}{ 89}$ \\  
\hline
 0.1570-0.1560 & $0.0483\errr{ 15}{ 13}$  & $0.1156\errr{ 29}{ 31}$  &
 $0.1937\errr{ 62}{ 66}$  & $0.4214\errr{ 89}{ 86}$ \\  
\hline
 0.1560-0.1560 & $0.0534\errr{ 16}{ 15}$  & $0.1199\errr{ 32}{ 31}$  &
 $0.2030\errr{ 63}{ 68}$  & $0.4103\errr{ 89}{ 90}$ \\  
\hline
\hline
\end{tabular}
\end{center}
\end{table}

\begin{table}
\begin{center}
\caption{\label{raw_decay_1565}Lattice results pseudoscalar and vector
meson decay constants at $\kappa_{\rm sea} = 0.1565$. For all
fits we find $1 \leq \chi^2/\mbox{d.o.f} \leq 2$.}
\begin{tabular}{|ccccc|}
\hline 
\hline
\multicolumn{5}{|c|}{$\kappa_{\rm sea} = 0.1565$}\\\hline
\multicolumn{5}{|c|}{nconfigs $ = 198$, nboot $= 200$, correlated}\\\hline
$(t_{\rm min}$,$t_{\rm max})$ & (9,15) &  & (10,15) &  \\
\hline 
\hline
$\kappa_1$-$\kappa_2$ & $\langle 0|A_0^l|\pi \rangle$ & $f_{\pi}/Z_A$
 & $3^{-1/2}\langle 0| V^l| V\rangle$ & $1/(Z_V f_{\rho})$ \\
\\
\hline
 0.1580-0.1580 & $0.0317\errr{ 19}{ 17}$  & $0.1024\errr{ 50}{ 43}$ &
 $0.162\errr{12}{11}$ & $0.481\errr{17}{18}$ \\
\hline
 0.1580-0.1575 & $0.0341\errr{ 20}{ 18}$  & $0.1047\errr{ 48}{ 43}$ &
 $0.165\errr{11}{11}$ & $0.472\errr{17}{18}$ \\ 
\hline 
 0.1575-0.1575 & $0.0365\errr{ 21}{ 17}$  & $0.1070\errr{ 50}{ 41}$ &
 $0.168\errr{11}{11}$ & $0.463\errr{17}{18}$ \\ 
\hline
 0.1580-0.1570 & $0.0365\errr{ 21}{ 17}$  & $0.1068\errr{ 50}{ 42}$ &
 $0.169\errr{11}{11}$ & $0.462\errr{16}{18}$ \\  
\hline
 0.1575-0.1570 & $0.0390\errr{ 22}{ 18}$  & $0.1091\errr{ 49}{ 40}$ &
 $0.172\errr{12}{11}$ & $0.454\errr{16}{17}$ \\  
\hline
 0.1580-0.1565 & $0.0389\errr{ 22}{ 18}$  & $0.1087\errr{ 50}{ 41}$ &
 $0.172\errr{12}{11}$ & $0.453\errr{16}{17}$ \\  
\hline
 0.1570-0.1570 & $0.0414\errr{ 23}{ 19}$  & $0.1113\errr{ 48}{ 41}$ &
 $0.176\errr{11}{11}$ & $0.446\errr{15}{17}$ \\  
\hline
 0.1575-0.1565 & $0.0414\errr{ 23}{ 19}$  & $0.1111\errr{ 48}{ 41}$ &
 $0.176\errr{11}{11}$ & $0.445\errr{15}{17}$ \\  
\hline
 0.1570-0.1565 & $0.0439\errr{ 22}{ 19}$  & $0.1133\errr{ 48}{ 39}$ &
 $0.180\errr{9}{9}$ & $0.438\errr{14}{16}$ \\  
\hline
 0.1580-0.1560 & $0.0412\errr{ 22}{ 18}$  & $0.1104\errr{ 50}{ 41}$ &
 $0.176\errr{12}{11}$ & $0.444\errr{15}{17}$ \\  
\hline
 0.1565-0.1565 & $0.0463\errr{ 21}{ 20}$  & $0.1153\errr{ 47}{ 39}$ &
 $0.184\errr{9}{9}$ & $0.431\errr{14}{15}$ \\ 
\hline
 0.1575-0.1560 & $0.0437\errr{ 23}{ 19}$  & $0.1129\errr{ 49}{ 40}$ &
 $0.180\errr{11}{10}$ & $0.437\errr{14}{16}$ \\  
\hline
 0.1570-0.1560 & $0.0462\errr{ 22}{ 19}$  & $0.1151\errr{ 48}{ 38}$ &
 $0.184\errr{9}{10}$ & $0.430\errr{13}{15}$ \\  
\hline
 0.1565-0.1560 & $0.0487\errr{ 22}{ 21}$  & $0.1172\errr{ 47}{ 41}$ &
 $0.188\errr{9}{10}$ & $0.424\errr{12}{15}$ \\  
\hline
 0.1560-0.1560 & $0.0512\errr{ 25}{ 21}$  & $0.1191\errr{ 47}{ 41}$ &
 $0.193\errr{9}{9}$ & $0.417\errr{12}{14}$ \\
\hline
\hline
\end{tabular}
\end{center}
\end{table}

\begin{table}
\begin{center}
\caption{\label{raw_decay_157}Lattice results pseudoscalar and vector
meson decay constants at $\kappa_{\rm sea} = 0.157$. For all
fits we find $1 \leq \chi^2/\mbox{d.o.f} \leq 2$.}
\begin{tabular}{|ccccc|}
\hline 
\hline
\multicolumn{5}{|c|}{$\kappa_{\rm sea} = 0.157$}\\\hline
\multicolumn{5}{|c|}{nconfigs $ = 198$, nboot $= 200$, correlated}\\\hline
$(t_{\rm min}$,$t_{\rm max})$ & (10,15) &  & (9,15) &  \\
\hline 
\hline
$\kappa_1$-$\kappa_2$ & $\langle 0|A_0^l|\pi \rangle$ & $f_{\pi}/Z_A$
 & $3^{-1/2}\langle 0| V^l| V\rangle$ & $1/(Z_V f_{\rho})$ \\
\\
\hline
 0.1575-0.1575 & $0.0302\errr{10}{ 9}$ & $0.0955\errr{ 27}{ 24}$ &
 $0.1491\errr{ 65}{ 66}$ & $0.4454\errr{ 79}{ 72}$ \\ 
\hline
 0.1575-0.1570 & $0.0326\errr{ 10}{9}$ & $0.0979\errr{ 26}{ 24}$ &
 $0.1538\errr{ 64}{ 70}$ & $0.4388\errr{ 79}{ 75}$ \\  
\hline 
 0.1570-0.1570 & $0.0350\errr{ 11}{10}$ & $0.1003\errr{ 26}{ 23}$ &
 $0.1586\errr{ 62}{ 71}$ & $0.4326\errr{ 75}{ 78}$ \\ 
\hline
 0.1575-0.1565 & $0.0349\errr{ 11}{10}$ & $0.1001\errr{ 26}{ 23}$ &
 $0.1584\errr{ 63}{ 72}$ & $0.4317\errr{ 75}{ 78}$ \\  
\hline
 0.1570-0.1565 & $0.0374\errr{ 12}{ 10}$ & $0.1026\errr{ 25}{ 25}$ &
 $0.1633\errr{ 67}{ 70}$ & $0.4261\errr{ 72}{ 75}$ \\  
\hline
 0.1565-0.1565 & $0.0398\errr{ 13}{ 11}$ & $0.1050\errr{ 27}{ 27}$ &
 $0.1680\errr{ 63}{ 66}$ & $0.4201\errr{ 73}{ 73}$ \\  
\hline
 0.1575-0.1555 & $0.0395\errr{ 12}{ 10}$ & $0.1040\errr{ 27}{ 25}$ &
 $0.1671\errr{ 67}{ 67}$ & $0.4167\errr{ 72}{ 73}$ \\  
\hline
 0.1570-0.1555 & $0.0420\errr{ 13}{ 12}$ & $0.1067\errr{ 28}{ 28}$ &
 $0.1721\errr{ 66}{ 67}$ & $0.4121\errr{ 72}{ 69}$ \\  
\hline
 0.1565-0.1555 & $0.0445\errr{ 14}{ 12}$ & $0.1091\errr{ 28}{ 28}$ &
 $0.1770\errr{ 62}{ 64}$ & $0.4071\errr{ 70}{ 67}$ \\  
\hline
 0.1555-0.1555 & $0.0493\errr{ 14}{ 13}$ & $0.1133\errr{ 26}{ 28}$ &
 $0.1863\errr{ 63}{ 63}$ & $0.3961\errr{ 68}{ 64}$ \\  
\hline
\hline
\end{tabular}
\end{center}
\end{table}

\begin{table}
\begin{center}
\caption{\label{raw_decay_1575}Lattice results pseudoscalar and vector
meson decay constants at $\kappa_{\rm sea} = 0.1575$. For all
fits we find $0.5 \leq \chi^2/\mbox{d.o.f} \leq 1$.}
\begin{tabular}{|ccccc|}
\hline 
\hline
\multicolumn{5}{|c|}{$\kappa_{\rm sea} = 0.1575$}\\\hline
\multicolumn{5}{|c|}{nconfigs $ = 198$, nboot $= 200$, correlated}\\\hline
$(t_{\rm min}$,$t_{\rm max})$ & (9,15) &  & (10,15) &  \\
\hline 
\hline
$\kappa_1$-$\kappa_2$ & $\langle 0|A_0^l|\pi \rangle$ & $f_{\pi}/Z_A$
 & $3^{-1/2}\langle 0| V^l| V\rangle$ & $1/(Z_V f_{\rho})$ \\
\\
\hline
 0.1575-0.1575 & $0.0251\errr{8}{9}$ & $0.0894\errr{ 26}{ 28}$ &
 $0.1286\errr{ 55}{ 61}$ & $0.4443\errr{99}{99}$ \\ 
\hline
 0.1575-0.1570 & $0.0276\errr{8}{9}$ & $0.0924\errr{ 26}{ 27}$ &
 $0.1321\errr{ 57}{ 60}$ & $0.4370\errr{99}{ 91}$ \\  
\hline 
 0.1570-0.1570 & $0.0301\errr{8}{9}$ & $0.0953\errr{ 26}{ 26}$ &
 $0.1361\errr{ 55}{ 55}$ & $0.4305\errr{ 96}{ 86}$ \\
\hline
 0.1575-0.1565 & $0.0300\errr{8}{9}$ & $0.0951\errr{ 27}{ 26}$ &
 $0.1360\errr{ 53}{ 57}$ & $0.4295\errr{ 95}{ 87}$ \\ 
\hline
 0.1570-0.1565 & $0.0325\errr{8}{9}$ & $0.0979\errr{ 24}{ 26}$ &
 $0.1403\errr{ 53}{ 57}$ & $0.4239\errr{ 95}{ 82}$ \\  
\hline
 0.1565-0.1565 & $0.0350\errr{8}{9}$ & $0.1004\errr{ 23}{ 25}$ &
 $0.1448\errr{ 53}{ 55}$ & $0.4180\errr{ 91}{ 84}$ \\  
\hline
 0.1575-0.1555 & $0.0347\errr{9}{10}$ & $0.0994\errr{ 24}{ 27}$ &
 $0.1439\errr{ 51}{ 57}$ & $0.4141\errr{ 91}{ 84}$ \\ 
\hline
 0.1570-0.1555 & $0.0372\errr{8}{9}$ & $0.1021\errr{ 23}{ 24}$ &
 $0.1487\errr{ 52}{ 56}$ & $0.4099\errr{ 90}{ 83}$ \\ 
\hline
 0.1565-0.1555 & $0.0396\errr{8}{10}$ & $0.1046\errr{ 23}{ 22}$ &
 $0.1535\errr{ 52}{ 58}$ & $0.4052\errr{ 88}{ 85}$ \\ 
\hline
 0.1555-0.1555 & $0.0444\errr{10}{10}$ & $0.1089\errr{ 21}{ 22}$ &
 $0.1629\errr{ 47}{ 57}$ & $0.3948\errr{ 83}{ 80}$ \\ 
\hline
\hline
\end{tabular}
\end{center}
\end{table}

%
%
\begin{table}
\begin{center}
\caption{Fit results for pseudoscalar, vector, nucleon and delta
  particles (in lattice units) from  ``symmetric'' fits.\label{table_pr}} 
\hspace{0.2cm}
\begin{tabular}{|c|cccc|}
particle & $a$  & $b$ & &  $\chisq$ \\ \hline 
PS & $-12.407(380)$ & $1.9666(540)$ & & $1.2/2$  \\
\hline \hline
particle & $m^{crit}$  & $c$ & $e$ &  $\chisq$ \\ \hline
V  & $0.3300(90)$  & $4.070(250)$  & 0 & $1.5/2$ \\ 
V  & $0.2928(410)$  & $6.44(2.40)$ & $-33.28(34.1)$ & $0.27/1$ \\ \hline
N  & $0.5012(190)$  & $6.960(460)$ & 0 & $2.7/2$ \\ 
N  & $0.4246(750)$  & $11.84(4.8)$ & $-67.99(69.1)$ &$1.5/1$ \\ \hline
$\Delta$  & $0.5851(240)$  & $6.482(670)$ & 0 & $6/2$ \\
$\Delta$  & $0.444(110)$   & $15.16(6.3)$ & $-118.9(82.0)$ & $3.5/1$ \\ \hline
$f_{\pi}$ & $0.0496(34)$  & $0.888(101)$  & 0 & $1.3/2$ \\ 
$f_{\pi}$ & 0.0423(137) & 1.354(778) & -6.58(11.15) & 0.89/1 \\ \hline
$1/f_{\rho}$ & $0.302(110)$  & $-0.372(290)$ & 0 & $0.37/2$ \\
$1/f_{\rho}$ & 0.294(43) & 0.170(2.204) & -7.45(31.44) & 0.31/1 \\
\end{tabular}
\end{center}
\end{table}
%
%
\begin{table}
\begin{center}
\caption{\label{alpha_value}Values of the
 coupling constant. We do not take into account
  the light-quark dependence of the plaquette in the determination of
  the strong coupling constant.}
\hspace{0.2cm}
\begin{tabular}{|c|c|ccc|}
  & $S_{\Box}$ & $\alpha_V(3.41/a)$ &  $\alpha_{\overline{MS}}(\pi/a)$ &
 $\alpha_{\overline{MS}}(1/a)$\\ 
\hline \hline
$\beta=5.6,N_F=2$ & (0.43012,0.42927,0.42837,0.42749)& 0.167 & 0.150 & 0.215 \\
\hline
$\beta=6.0,N_F=0$ & 0.406318                 & 0.152 & 0.138 & 0.205 \\ 
\end{tabular}
\end{center}
\end{table}
%
%
%
%
\begin{table}
\begin{center}
\caption{Fit results for the masses of pseudoscalar and
vector particles in the strange sector, according to
eqs. \ref{quarkmatrix}, \ref{massfit_generic}
and \ref{thorsten} \label{s_chiral}} 
\hspace{0.2cm}
\begin{tabular}{|ccccc|}   
PS(linear fit) & & $a'$  & $b'$ &  $\chisq$ \\ \hline
 & - & $3.93(12)$ & $1,01(11)$ & $26/31$  \\
\hline \hline
V(linear fit) &  $m'^{\sf crit}$  & $c'$  & $d'$ &  $\chisq$ \\ \hline
 & $0.3300(93)$ & $4.07(25)$  & $0.948(31)$  & $13/31$ \\ \hline \hline
V(quadr. fit) &  $m'^{\sf crit}$  & $c'$  & $d'$ &  $\chisq$ \\ \hline
 & $0.2928(412)$ & $6.44(2.4)$  & $0.908(145)$  & $3/29$ \\ \hline
 &  $e'$  & $f'$  & $g'$ & \\ \hline
 & $-33.28(34.3)$ & $-2.19(4.79)$ & $2.25(2.03)$ & \\ \hline
\end{tabular}
\end{center}
\end{table}
%
%
\begin{table}[tb]
\begin{center}
\caption{Collection of results for $\kst$.\label{strangefits}}
\hspace{0.2cm}
\begin{tabular}{|ccc|}
particle &  $\kappa^{\rm strange}_{lin}$ &  $\kappa^{\rm strange}_{quad}$ \\
\hline
$K$ & $0.15654(11)$ & $0.15694(42)$ \\
\hline
$K^*$ & $0.15561(14)$ & $0.15590(57)$ \\
\hline
$\phi$ & $0.15563(14)$ & $0.15598(50)$ \\
\end{tabular}
\end{center}
\end{table} 
%
%
\begin{table}
\begin{center}
\caption{Lattice results in the strange quark sector.\label{fixedseastrange}}
\hspace{0.2cm}
\begin{tabular}{|ccc|}
observable & lin fit to  $m_{\rho}$ &  quad fit to $m_{\rho}$ \\ \hline
$m_{K}(\phi)$    & $0.259(6)$ & $0.241(25)$ \\
$m_{\phi}(\phi)$ & $0.443(12)$ & $0.395(54)$ \\
$m_{K^*}(\phi)$  & $0.388(11)$ & $0.345(47)$ \\ \hline
$m_{K}(K^*)$    & $0.259(6)$ & $0.245(28)$ \\
$m_{\phi}(K^*)$ & $0.443(12)$ & $0.398(56)$ \\
$m_{K^*}(K^*)$  & $0.288(11)$ & $0.346(48)$ \\ \hline
$m_{K(K)}$    & $0.215(6)$ & $0.192(26)$ \\
$m_{\phi}(K)$ & $0.407(13)$ & $0.356(53)$ \\
$m_{K^*}(K)$  & $0.371(11)$ & $0.326(47)$ \\ \hline
$f_{K}(\phi)$      & 0.0633(33) & 0.0615(38)   \\
$1/f_{\phi}(\phi)$ & 0.2742(93) & 0.2777(97)   \\ \hline
$f_{K}(K^*)$      &  0.0633(33) & 0.0619(38)   \\
$1/f_{\phi}(K^*)$ &  0.2741(93) & 0.2769(104) \\ \hline
$f_{K}(K)$      &    0.0591(33) & 0.0572(34)   \\
$1/f_{\phi}(K)$ &    0.2832(97) & 0.2871(105)   \\
\end{tabular}
\end{center}
\end{table}
%
%
\begin{table}
\begin{center}
\caption{\label{table_latt}Values of the inverse lattice spacing
  obtained from different   observables (at the light quark mass).
The lattice value of $m_N$ results from a quadratic extrapolation,
the value of $f_\pi$ stems from a linear extrapolation.}
\hspace{0.2cm}
\begin{tabular}{|ccc|}
observable & $m(\ksl)$ & $a^{-1}$ \\ \hline
$m_\rho$ (linear fit)    & $0.334(9)$    & $2.30(6)$    \\ \hline
$m_\rho$ (quadratic fit) & $0.297(41)$   & $2.58(37)$    \\ \hline
$m_N$ (rho linear)       & $0.435(72)$  & $2.16(40)$  \\ \hline
$f_\pi$ (rho linear)     & $0.0505(34)$ & $2.62(18)$  \\ 
\end{tabular}
\end{center}
\end{table}
%
%
\begin{table}
\begin{center}
\caption{Physical results in the light quark sector.\label{fixedsea1}}
\hspace{0.2cm}
\begin{tabular}{|cccc|}
 $m_N[GeV]$ & $m_\Delta[GeV]$ & $f_{\pi}[GeV]$ & $1/f_\rho$ \\ \hline
 linear vector & & & \\ \hline
 1.00(16) & 1.05(23) & 0.116(8) & 0.302(11) \\ \hline
 quadr. vector & & & \\ \hline
 1.12(25) & 1.17(28) & 0.130(21) & 0.302(11) \\ \hline
\end{tabular}
\end{center}
\end{table}
%
%
\begin{table}
\begin{center}
\caption{Physical results for masses in the strange quark sector.\label{fixedsea2}}
\hspace{0.2cm}
\begin{tabular}{|cccccc|}
 $m_K(K^*)[GeV]$ & $m_K(\phi)[GeV]$ &  $m_{K^*}(K)[GeV]$ & $m_{K^*}(\phi)[GeV]$ & $m_{\phi}(K)[GeV]$ &
 $m_{\phi}(K^*)[GeV]$ \\ \hline
 linear vector & & & & & \\ \hline
 0.596(9) & 0.595(9) & 0.853(3) & 0.894 & 0.937(6) & 1.020 \\ \hline
 quadr. vector & & & & & \\ \hline
 0.633(47) & 0.623(44) & 0.842(12) & 0.890(4) & 0.918(24) & 1.029(9) \\ \hline
 \end{tabular}
\end{center}
\end{table}
%
%
\begin{table}
\begin{center}
\caption{Physical results for decay constants in the strange quark sector.\label{fixedsea3}}
\hspace{0.2cm}
\begin{tabular}{|cccccc|}
 $f_K(\phi)[GeV]$ & $f_K(K^*)[GeV]$ &  $f_K(K)[GeV]$  & $1/f_{\phi}(\phi)$ &
 $1/f_{\phi}(K^*)$ & $1/f_{\phi}(K)$ \\ \hline
 linear vector & & & & & \\ \hline
 0.1456(71) & 0.1457(71) & 0.1360(72) & 0.2742(93) & 0.2741(93) &
 0.2832(97)\\ \hline 
 \end{tabular}
\end{center}
\end{table}
%
%
\begin{table}
\begin{center}
\caption{Physical results in the quenched sector. Fits to nucleon and
$\Delta$ are quadratic, all other fits are linear.
To set $\kappa^{strange}$ we used the Kaon mass for $f_K$ and $m_{K^*}$,
and the $\phi$ mass for $1/f_{\phi}$, $m_K$ and $J$. \label{quenched}}
\hspace{0.2cm}
\begin{tabular}{|ccccc|}
 $m_N[GeV]$ & $m_\Delta[GeV]$ & $f_{\pi}[GeV]$ &  $1/f_{\rho}$ & \\ \hline
  1.061(67) & 1.301(63)       & 0.1325(48)     &   0.3271(52)  & \\
  \hline
 $m_K[GeV]$     &  $m_{K^*}[GeV]$   & $f_K[GeV]$ & $1/f_{\phi}$ & $J$
  \\
 \hline
   0.5590(64)  &     0.8647(23)     & 0.1518(46) & 0.2982(47) &  0.38(1) \\
 \hline
\end{tabular}
\end{center}
\end{table}
%
%

%
%


%
%
%
%
%
%
%
%
%
%
%
%
%
%
%
%
%
%
%
%

\begin{thebibliography}{99}

\bibitem{yoshie}{T.~Yoshi\'e, 
Nucl.Phys. {\bf B} (Proc.Suppl.) {\bf 63}(1998)3.}

\bibitem{weingarten}{F.~Butler et al, Nucl. Phys. {\bf B430} (1994)179.}

\bibitem{tsukuba}{CP-PACS collaboration, S. Aoki et al.,
Nucl.Phys. (Proc.Suppl.) {\bf 60A} (1998)14.} 


\bibitem{hmc}{S.~Duane, A.~Kennedy, B.~Pendleton, and
D.~Roweth, Phys. Lett. {\bf B195} (1987)216.}

\bibitem{ourmethods}{A.~Frommer, V.~Hannemann, Th.~Lippert,
B.~N\"ockel, and K.~Schilling, Int. J. Mod. Phys. {\bf C5}(1994)1073;
A.~Frommer, S.~G\"usken, Th.~Lippert, B.~N\"ockel, and K.~Schilling,
Int. J. Mod. Phys. {\bf C6}(1995)627.}

\bibitem{ape}{
R. Tripiccione, Int.J.Mod.Phys. {\bf C4}(1993)425.} 

\bibitem{qmasses} {\sesam--Collaboration, N.~Eicker et al,
    Phys.~Lett.{\bf B}407 (1997)290.}

\bibitem{GOTTLIEB} S. Gottlieb, W. Liu, D. Toussaint, L. R. Renken, 
and R. L. Sugar, Phys. Rev. {\bf D 35} (1987) 2531.

\bibitem{ROSSI} T. DeGrand and P. Rossi, Comp. Phys. Comm. {\bf 60 }
(1990) 211.

\bibitem{FISCHER}S.~Fischer et al., Comp. Phys. Comm. 98 (1996)20;
Nucl. Phys. {\bf B} (Proc. Suppl.){\bf 53} (1997)990.

\bibitem{BROWER} R. C. Brower, T. Ivanenko, A. R. Levi,
K. N. Orginos, Nucl. Phys. {\bf B484} (1997) 353.

\bibitem{MACKENZIE}P. Mackenzie, Phys. Lett. {\bf 226B} (1989) 369.

\bibitem{SESauto} {\sesam--Collaboration, Th. Lippert et al.,
Performance of the Hybrid Monte Carlo for QCD with Wilson fermions; ,
in preparation.}

\bibitem{TOPOL}B. All\'es et al., IFUM 608-FT, IFUP-TH 10/98, 
 HLRZ-98.6, HUP-EP-98/12, WUP-TH 98-6, hep-lat/9803008.

\bibitem{Wupsmear} {S.~G\"usken et al., Nucl.~Phys. {\bf B}(Proc.~Suppl)
{\bf 17}(1990), 361.} 
%



\bibitem{PRD} {R.M. Barnett et al., Physical Review {\bf D54}(1996)1
    (Paticle Data Booklet). We use the following masses:
    $m_{\pi} = {1 \over 2} (M_{\pi^{\pm}} + M_{\pi^{0}}) = 137.3\;{\sf
      MeV}$; $M_{\rho} = 769\;{\sf MeV}$; $M_{\phi} = 1.019.4\;{\sf
      GeV}$; $M_{K} = {1\over 2} (M_{K^{\pm}} + M_{K^{0}}) =
    495.68\;{\sf MeV}$; $M_{K^*} = 892\;{\sf MeV}$.}

 \bibitem{Jpar}
  {P.~Lacock and C.~Michael, Phys.Rev.~ {\bf D 52}(1995)5213.}

\bibitem{chiral_pbt}E.~Jenkins, A.V.~Manohar, M.B.~Wise, Phys. Rev.
  Lett. {\bf 75} (1995)2272.

\bibitem{tchl} {T$\chi$L - Collaboration, L.~Conti, L.~Giusti,
    U.~Gl\"assner, S.~G\"usken, H.~Hoeber, P.~Lacock, Th.~Lippert,
    G.~Martinelli,
    F.~Rapuano, G.~Ritzenh\"ofer, K.~Schilling, G.~Siegert and
    A.~Spitz; in preparation.} 


\bibitem{lepage} {G.P.~ Lepage and P.B.~
    Mackenzie, Phys.~ Rev.~ {\bf D 48} (1993)2250.}


\end{thebibliography}
\end{document}